\documentclass[a4paper,12pt]{article}
\usepackage[utf8]{inputenc}
\usepackage{cancel}
\usepackage{ulem}
\usepackage{amsfonts}
\usepackage{amssymb}
\usepackage{graphicx}
\usepackage{amsmath}
\usepackage{enumerate}
\usepackage{mathtools}
\usepackage{subfig}
\usepackage{color}
\usepackage{float}
\usepackage{here}
\usepackage{cite}
\usepackage{mathrsfs}
\usepackage{float,epsfig}
\usepackage{dcolumn}

\usepackage{graphicx}
\usepackage{bm}
\usepackage{amsmath,amssymb,amsthm}
\usepackage[colorlinks=true,linkcolor=blue,citecolor=red]{hyperref}
\makeatother

\newcommand{\be}{\begin{equation}}
\newcommand{\ee}{\end{equation}}
\newcommand{\bea}{\begin{eqnarray}}
\newcommand{\eea}{\end{eqnarray}}
\numberwithin{equation}{section}
\begin{document}
\title{\normalsize
\phantom{fff}

\vspace{-1cm}
{\bf \Large   Deflection angle and shadow of slowly rotating \\ \vspace{0.2cm}  black holes in galactic nuclei}}
\author{ A. El Balali $^{1}$\thanks{anas.elbalali@gmail.com},  M. Benali $^{1}$ \thanks{mohamed\_benali4@um5.ac.ma},  M. Oualaid $^{1}$ \thanks{mohamed\_oualaid@um5.ac.ma}
\hspace*{-4pt} \\
{\small $^1$
Département de Physique, Equipe des Sciences de la matière et du rayonnement, ESMAR
}\\{\small Faculté des Sciences, Université Mohammed V de Rabat, Rabat, Morocco }\\
} \maketitle

	\begin{abstract}
		{\noindent}
In this paper, we construct the slowly rotating case of an asymptotically flat supermassive black hole embedded in dark matter using Newman-Janis procedure. Our analysis is carried with respect to the involved parameters including the halo total mass $M$ and the galaxy's lengthscale $a_0$. Concretly, we investigate the dark matter impact on the effective potential and the photon sphere. In particular, we find that the lengthscale $a_0$ controles such potential values. Indeed, for low $a_0$ values, we find that the halo total mass $M$ decreases the potential values significantly while for high $a_0$ values such impact is diluted. Regarding the shadow aspects, we show that the shadow size is much smaller for high values of $a_0$ while the opposite effect is observed when the halo total mass $M$ is increased. By comparing our case to the slowly rotating case, we notice that the former exhibits a shadow shifted from its center to the left side. Finally, we compute the deflection angle in the weak-limit approximation and inspect the dark matter parameters influence. By ploting such quantity, we observe that one should expect lower bending angle values for black holes in galactic nuclei.

	\end{abstract}

\newpage

\tableofcontents

\newpage

\section{Introduction}
Lately, Einstein theory of gravity has been at the center of interest due to its fascinating prediction of black holes \cite{O0}. To uncover more about their true nature, investigations have been led in different directions and for many black hole types \cite{O1,O2,O3,O4,O5,O6,O6-1,O6-2,O6-3,O6-4,O7,O8,O9,O10}. These objects in such a theory, and many other gravity theories, are characterized by an extremly intense gravity yielding their hardly observational aspect. However, the Event Horizon Telescope have achieved a breakthrough by anouncing their first detection of a supermassive black hole at the center of $M87$ elliptical galaxy in $2019$ \cite{I1}. Recently, they have provided a second image describing the influence of a magnetic field on the black hole shadow and accretion disk \cite{I2,I3}. Such images describe a black hole illuminated by external sources showing a dark spot, associated to the black hole shadow, together with its accretion disk. In general, the shadow of a non rotating black hole is a strandard circle while the rotating one exhibit a D-shaped shadow caused by the spacetime dragging effects. Therefore, the advances in black holes observations  have peaked the interest of physicist from all around the world making the investigation of black hole's shadow, accretion disk and deflection angle increase drastically. In fact, the agreement between the Kerr black hole shadow and EHT images have motivated futher inspections of other black hole's optical aspects, and in differents backgrounds, that may probably match futur observations. Indeed, many researches have studied the shadow of Schwarzschild \cite{I4} , Kerr \cite{I5, I6}, Kaluza Klein \cite{I7}, naked singularities \cite{I8}, Weyl black holes \cite{I9} and many others \cite{I10, I11,I12,I13,I14,I15,A1,A2,A3,A4,A5,A6,A7,A8,A9}.
In the frame of General Relativiy, gravity is rather a spacetime curvature than a force. A straightforward consequence of this feature is that light rays are deflected when they propagate in a curved spacetime. Such a phenomenon, called gravitational lensing, is an important method with great impact on astronomy and cosmology. Two different categories in leterature could be distinguished when it comes to  gravitational lensing. First, the weak gravitational lensing have been used to compute astronomical objects mass's or to find the rapid univers expansion potential cause \cite{GL1,GL2,GL3}. Second, the strong gravitational lensing provides information about the black hole image position and time delay \cite{GL4,GL5,GL6,GL7,GL8,GL9,GL10}. Such applications have led to various investigations of light deflection by different black hole spacetimes \cite{GL11,GL12,GL13,GL14,GL15,GL16,GL17,GL18,GL19,GL20,GL21,GL22,GL23,GL24,GL25}.

On the other hand, it is believed that a supermassive black hole resides in many galaxy's center. Since $85 \%$ of the universe consists of an invesible dark matter, its only natural to carry inspections of black holes immersed in this astrophysical environement \cite{I16,I17,I18,I19,I20,I21,I22}. Toward this aim, many studies have been elaborated both in the presence of dark matter and dark energy \cite{I30,I31,I32}. Although a direct dark matter detection has not yet been obtained, strong observational evidence of its existance in giant elliptical and spiral galaxies has been provided \cite{I34}. Besides, indications of elliptical and spiral galaxies being embedded in a giant dark matter  halo have been observed using astrophysical techniques \cite{I36,I37,I38,I39,I40}. To gain further insight about these situations, theoritical advances have been carried. For instance, the circular geodesics of rotating black hole with quintessence has been computed in the presence of an external magnetic field \cite{I41,I42}. The cold dark matter surronding a black hole in a phontom field has been calculated in \cite{I43}. Shadow of non rotating and rotating black hole in perfect fluid dark matter has been studied in \cite{I31}. For the rotating charged case in perfect fluid dark matter, it has been investigated in \cite{I46}. The elaboration of the weak deflection angle through dark matter by black holes and wormholes using Gauss-Bonnet theorem has been done in  \cite{II1,II2}. In papers \cite{II3,II4}, the weak deflection angle by a rotating black hole surrounded by dark matter  has been determined.
 However, many of the research have been relying on Newtonian approaches to these dark matter configurations \cite{I47,I48,I49,I50,I51}. More recently, an exact "fluid-hairy" black hole solution describing a realistic dark matter distribution has been obtained \cite{Car1,Car2,Car3}. This interesting solution has the advantage to follow a Sersic density profile linked to the Hernquis model which has been observationally confirmed in elliptical galaxies \cite{I52,I53}. In this way, we could consider that the provided metric describes a black hole in active galactic nuclei. 

In this paper, the main goal is to contribute to this activities by studying slowly rotating supermassive black holes in active galactic nuclei.  Concretely, we investigate the dark matter impact on the supermassive black hole geometry. Such impact lead to perturbations of the null geodesics and photon orbits. In particular, we determine and analyze the shadow behaviors as a function of the involved parameters and compare them to the slowly rotating and Schwarzschild case. Indeed, such illustrations show that the frame dragging effect in the presence of dark matter configuration is opposed to the ordinary slowly rotating case. Then, the deflection of light is inspect with the use of Gauss-Bonnet theorem in the weak-limit approximation.

This work is organized as follows:  In section \eqref{II}, we briefly review the non rotating fluid-hairy black hole solution. In section \eqref{III}, we generate the slowly rotating solution through Newman-Janis algorithm. Section \eqref{IV} is dedicated to the elaboration of photon sphere, shadow aspects. Then, we visualize the $4U  1543-475$ and $GRO J1655-40$ black hole shadows in the considered dark matter configuration. Section \eqref{v}, concerns the study of the deflection angle which is obtained using the Gauss-Bonnet theorem. Finally, a conclusion and open questions are developed in section \eqref{VI}.
\section{Non rotating fluid-hairy black hole}
\label{II}
In the context of Einstein's gravity coupled minimally to an anisotropic fluid corresponding to dark matter, an analytical solution has been derived describing a non rotating black hole at the center of a Hernquist-type density distribution \cite{I52,I53}. These spacetimes can describe the geometry of supermassive black holes at galactic nuclei. The associated geometry is represented by the following metric
\begin{equation}
ds^2=-f(r) dt^2 + \frac{dr^2}{g(r)}+r^2 \left( d\theta^2 + \sin^2\theta  d\phi^2 \right).
\label{first}
\end{equation}
The metric functions $f(r)$ and $g(r)$ are given by
\begin{align}
f(r)&=\left(1-\frac{2M_{BH}}{r}  \right) e^{-\pi\sqrt{\frac{M}{2a_0-M+4M_{BH}}}+2\sqrt{\frac{M}{2a_0-M+4M_{BH}}}\arctan\left( \frac{r+a_0-M}{\sqrt{M\left( 2a_0-M+4M_{BH} \right)}} \right)}, \\
g(r)&=1-\frac{2m(r)}{r}=1-\frac{2M_{BH}}{r}-\frac{2Mr}{\left(a_0+r \right)^2} \left( 1-\frac{2M_{BH}}{r} \right)^2,
\end{align}
where $M_{BH}$ represents the black hole mass, $M$ is associated to the "halo" total mass, and $a_0$ is a typical length scale. The black hole solution described by the metric \eqref{first}, correspond to the energy density distribution 
\begin{equation}
\rho=\frac{m'}{4\pi r^2}=\frac{M \left( a_0+2M_{BH} \right)\left(1-2M_{BH}/r  \right)}{2 \pi r\left( r+a_0 \right)^3}.
\label{d}
\end{equation}
When $M_{BH} \to 0$ , it is easy to check that we recover Hernquist-type density $\rho=\frac{M a_0}{2 \pi r \left(r+a_0  \right)^3}$ which is observationally confirmed in elliptical galaxies. The density distribution \eqref{d}, has a maximum located at 
\begin{equation}
r_M=\frac{1}{8} \left(\sqrt{a_0^2+44 \, a_0 \, M_{BH}+100 M_{BH}^2}-a_0+10 \, M_{BH}\right),
\end{equation}
It is observed that such quantity does not depend on the halo dark matter total mass $M$. Besides, one notices that $r_M$ increases (decreases) when the black hole mass $M_{BH}$ increase (decrease).
Such maximum corresponds to the following density
\begin{align}
\rho_M = & \frac{2048 M (a_0+2 M_{BH}) }{\pi  \left(a_0-10 M_{BH} -\sqrt{a_0^2+44 a_0 M_{BH}+100 M_{BH}^2}\right)^2} \nonumber \\
& \times \frac{\left(a_0+6 M_{BH} -\sqrt{a_0^2+44 a_0 M_{BH}+100 M_{BH}^2}\right)}{ \left(7 a_0+10 M_{BH} +\sqrt{a_0^2+44 a_0 M_{BH}+100 M_{BH}^2}\right)^3}.
\end{align}
From such equation, one remarks that the density $\rho_M$ decreases when the black hole mass $M_{BH}$ increases. 

At the geometrical level, the black hole horizon is located at $r=2M_{BH}$ while the curvature singularity is at $r=0$. Besides, one also finds a curvature singularity at $r=M-a_0 \pm \sqrt{M^2-2Ma_0-4M M_{BH}}$. However, such singularity does not follow the astrophysical configuration since $M>2\left(a_0+2M_{BH}\right)$. For a realistic solution, one has to assume the inequalities $M_{BH}<<M<<a_0$. 
It is worth noting that the Schwarzschild black hole geometry is recovered by taking the limit $M \to 0$ in the metric \eqref{first}.
\section{Slowly rotating black hole solution}
\label{III}
In this section, we construct the slowly rotating solution of a black hole in galactic nuclei. At the same time, we give a brief review of the Newman-Janis procedure for a general static and spherically symmetric metric \cite{III1}. In Boyer Lindquist coordinates $\left(t,r,\theta, \phi \right)$, such a metric can be written as
\begin{equation}
ds^2=-f(r)dt^2+g(r)^{-1}dr^2+h(r)d\Omega^2,
\label{Met1}
\end{equation}
where $d\Omega^2=d\theta^2+\sin^2 \theta d\phi^2$ and $h(r)=r^2$. The Newman-Janis method is used to construct a stationary, axially symmetric and rotating solution from a non rotating one. Firstly,  to generate such a solution we  rewrite the metric \eqref{Met1} in Eddington-Finkelstein coordinates $\left(u,r,\theta, \phi \right)$ using the following transformation
\begin{equation}
du=dt-\frac{dr}{\sqrt{f(r) g(r)}}.
\end{equation}
As a result, the metric \eqref{Met1} in the advanced Eddington-Finkelstein coordinates is given by
\begin{equation}
ds^2=-f(r)du^2-2\sqrt{\frac{f(r)}{g(r)}}dudr+h(r)d\Omega^2,
\label{Met2}
\end{equation}
Secondly, the nonzero components of the resulting inverse metric can be introduced using the null tetrad $\left( l^\mu,n^\mu,m^\mu,\overline{m}^\mu \right)$ as
\begin{equation}
g^{\mu \nu}=-l^\mu n^\nu -l^\nu n^\mu+m^\mu \overline{m}^\nu +m^\nu \overline{m}^\mu,
\end{equation}
with
\begin{align}
l^\mu &=\delta^\mu_r, \\
n^\mu &= \sqrt{\frac{g(r)}{f(r)}} \delta^\mu_u-\frac{g(r)}{2} \delta^\mu_r, \\
m^\mu &= \frac{1}{\sqrt{2 h(r)}} \left(\delta^\mu_\theta + \frac{i}{\sin \theta} \delta^\mu_\phi  \right),\\
\overline{m}^\mu &= \frac{1}{\sqrt{2 h(r)}} \left(\delta^\mu_\theta - \frac{i}{\sin \theta} \delta^\mu_\phi  \right).
\end{align}
The over line is associated to complex conjugation, and the null tetrad satisfy the following equations
\begin{align}
l_\mu l^\mu &=n_\mu n^\mu=m_\mu m^\mu=l_\mu m^\mu=n_\mu m^\mu=0, \\
l_\mu n^\mu &=-m_\mu \overline{m}^\mu=-1.
\end{align}
Thirdly, we need to complexify the coordinate system as follows
\begin{equation}
x^{\prime \mu}=x^{ \mu}+i a \left( \delta^\mu_r-\delta^\mu_u \right) \cos \theta,
\label{comp}
\end{equation}
where $a$ is the spin parameter. Accordingly, the functions $\left\lbrace f(r),g(r),h(r)  \right\rbrace$ tranform to \\ $\left\lbrace F(r^{\prime}),G(r^{\prime}),H(r^{\prime})  \right\rbrace$ and $\delta^\mu_r \rightarrow \delta^\mu_r$, $\delta^\mu_u \rightarrow \delta^\mu_u$, $\delta^\mu_\theta \rightarrow \delta^\mu_\theta+ia \sin \theta \left(\delta_u^\mu-\delta_r^\mu \right)$, $\delta^\mu_\phi \rightarrow \delta^\mu_\phi$. In this way, the null tetrad are given by
\begin{align}
\label{Tet1}
l^{\prime \mu} &=\delta^\mu_r, \\
\label{Tet2}
n^{\prime \mu} &=\sqrt{\frac{G \left(r^\prime \right)}{F \left(r^\prime \right)}}-\frac{G \left(r^\prime \right)}{2} \delta^\mu_r, \\
\label{Tet3}
m^{\prime \mu} &=\frac{1}{\sqrt{2 H \left(r^\prime \right)}} \left(ia \sin \theta \left(\delta^\mu_u -\delta_r^\mu \right) +\delta^\mu_\theta +\frac{i}{\sin \theta} \delta_\phi^u \right).
\end{align}
It is worth noting that several ways of complexification can be found in literature. However, the complexification of transformation \eqref{comp} is known to generate rotating solutions succefully \cite{III2}. The inverse metric components are derived using the vectors \eqref{Tet1}, \eqref{Tet2} and \eqref{Tet3} which gives
\begin{align}
g^{uu} &=\frac{a^2 \sin^2 \theta}{H \left(r^\prime \right)}, \quad g^{u \phi}=\frac{a}{H \left(r^\prime \right)}, \\
g^{\phi \phi}&=\frac{1}{H \left(r^\prime \right) \sin^2 \theta}, \quad g^{\theta \theta}=\frac{1}{H \left(r^\prime \right)}, \\
g^{rr} &= G \left(r^\prime \right)+ \frac{a^2 \sin^2 \theta}{H \left(r^\prime \right)}, \quad g^{r \phi}=-\frac{a^2 \sin^2 \theta}{H \left(r^\prime \right)}, \\
g^{ur} &=-\sqrt{\frac{G \left(r^\prime \right)}{H \left(r^\prime \right)}}-\frac{a^2 \sin^2 \theta}{H \left(r^\prime \right)}.
\end{align}
The resulting metric in the advanced Eddington-Finkelstein coordinates is written as
\begin{align}
ds^2 = &-F\left(r'  \right)du^2-2\frac{F\left(r'  \right)}{G\left(r' \right)} du dr+2 a \sin^2 \theta \left(F\left(r'  \right)- \sqrt{\frac{F\left(r'  \right)}{G\left(r'  \right)} }  \right)du d\phi \nonumber \\
& + 2 a \sqrt{\frac{F\left(r' \right)}{G\left(r' \right)} } \sin^2\theta dr d\phi+ H(r')d\theta^2+\sin^2\theta \left[ H(r')+a^2 \sin^2\theta \left(2\sqrt{\frac{F(r')}{G(r')}} -F(r') \right) \right] d\phi^2.
\label{Met3}
\end{align}
The final step of Newman-Janis method consist of writing the metric in Boyer-Lindquist coordinates by perfoming the following transformations
\begin{equation}
du=dt^\prime+ A(r) dr, \quad d\phi=d\phi^\prime+B(r) dr.
\label{TRSF}
\end{equation}
Replacing these transformations in the metric \eqref{Met3} and taking $g_{tr}$ and $g_{r \phi}$ equal to zero,  we obtain
\begin{align}
\label{A}
A(r)& =-\frac{\sqrt{\frac{G \left(r, \theta \right)}{F \left(r, \theta \right)}} H\left(r, \theta \right)+a^2 \sin^2 \theta}{G\left(r, \theta \right) H\left(r, \theta \right) + a^2 \sin^2\theta}, \\
\label{B}
B(r)& =-\frac{a}{G\left(r, \theta \right) H\left(r, \theta \right) + a^2 \sin^2\theta}
\end{align}
It should be noted that the transformations \eqref{TRSF} are valid only if the left hand side of \eqref{A} and \eqref{B} are independent of $\theta$. In fact, the transformations \eqref{A}-\eqref{B} are not possible in general. However, an accurate transformation is achieved when the slow rotation limit $a^2 \to 0$ is considered. In this case, the metric functions $f(r), g(r)$ and $h(r)$ do not depend on $\theta$ after the complexification. Besides, we assume $a^2 << \sqrt{\frac{g(r)}{f(r)}} h(r)$ and  $a^2<< g(r)h(r)$ yielding an independent right hand side of equations \eqref{A} and \eqref{B}. Finally, we insert the equations \eqref{A} and \eqref{B} in \eqref{Met3} to obtain the slowly rotating spacetime metric
\begin{equation}
ds^2=-f(r) dt^2 + \frac{dr^2}{g(r)}+h(r) d\Omega^2-2a \,  e(r) \sin^2 \theta dt d\phi,
\label{fmetric}
\end{equation}
where $e(r)=\sqrt{\frac{f(r)}{g(r)}}-f(r)$. For such a slowly rotating solution, the energy momentum tensor should be determined. Indeed, such quantity is expressed as a function of the involved parameters in the appendix \eqref{app1}.
\section{Shadow aspects}
\label{IV}
In this section, we analyze the shadow behavior of the slowly rotating black hole in galactic nuclei described by the metric \eqref{fmetric}.
\subsection{Null geodesics and photon orbits}
\begin{figure}[h]
\begin{tabular}{ll}
\includegraphics[scale=0.55]{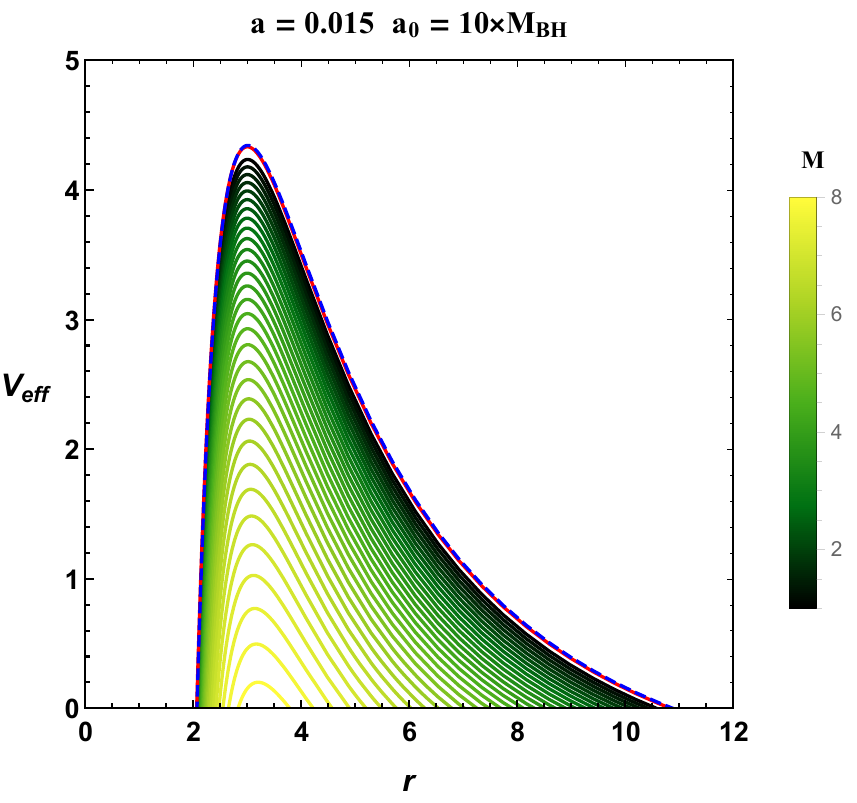} &\includegraphics[scale=0.55]{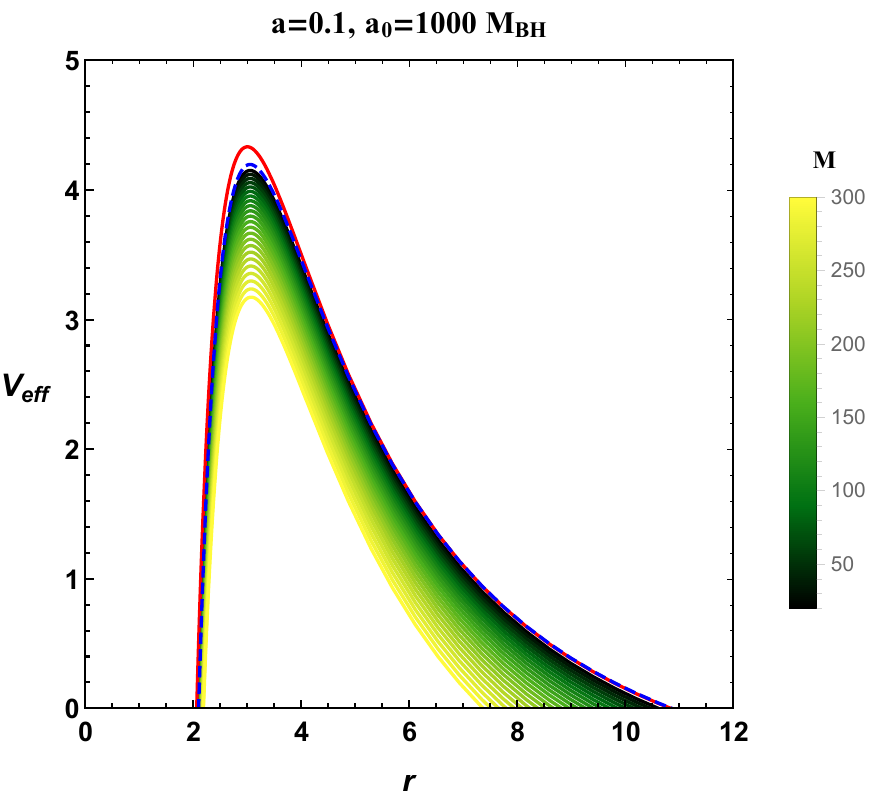}
\end{tabular}
\caption{ \footnotesize Effective potential for different values of the halo total mass $M$, length scale $a_0$,  the spin $a$ and fixed black hole mass $\left( M_{BH}=1 \right)$. The blue dashed curve is associated to the slowly-rotating Kerr black hole  with a spin $a=0.015$ (left) and $a=0.1$ (right) while the red curve correspond to Schwarzschild case. We take $E=\mathcal{K}=1$ and $L=12$.}
\label{V}
\end{figure}

To investigate the evolution of the photon around the considered black hole, one needs to derive the equation of motion \cite{1S1}. To do so, we exploit the following Hamilton-Jacobi equation 
\begin{equation}
\frac{\partial S}{\partial \tau}=-\frac{1}{2} g^{\mu \nu} \frac{\partial S}{\partial x^\mu} \frac{\partial S}{\partial x^\nu},
\label{HJ}
\end{equation}
where $\tau$ represents the affine parameter of the null geodesic and the Jacobi action can be separated in the following way
\begin{equation}
S=\frac{1}{2} m_0^2 \tau-E t + L \phi+S_r\left( r \right)+S_\theta \left( \theta \right).
\label{S}
\end{equation}
For the case of a photon, the mass $m_0$ is equal to zero. $E$ and $L$ are associated to the energy and angular momentum of the photon and the two functions $S_r\left( r \right)$ and $S_\theta \left( \theta \right)$ depend only on $r$ and $\theta$ respectively.
In the slow rotation regime, we obtain by replacing the Jacobi action \eqref{S} into the Hamilton-Jacobi equation \eqref{HJ} the following result
\begin{align}
0= & -f(r)^{-1} \left( \frac{\partial S}{\partial t} \right)^2+\frac{1}{r^2} \left( \frac{\partial S}{\partial \phi}  \right)^2 - 2a \sin^2 \theta \frac{e(r)}{r^2 f(r)}  \left( \frac{\partial S}{\partial t} \right) \left( \frac{\partial S}{\partial \phi}  \right) \\
& + g(r)  \left( \frac{d S_r}{d r} \right)^2 +\frac{1}{r^2 \sin^2 \theta}  \left( \frac{d S_\theta}{d \theta} \right)^2 +\mathcal{O}\left( a^2 \right).
\end{align}
Further calculations and simplifications provides
\begin{align}
r^4   g(r) \left( \frac{d S_r}{d r} \right)^2 &=\frac{E^2 r^4}{f(r)}+2a E L r^2 \frac{e(r)}{f(r)}-r^2 \left(L^2 + \mathcal{K} \right), \\
\frac{1}{\sin^2 \theta} \left( \frac{d S_\theta}{d \theta} \right)^2 & =\mathcal{K}-2a E L \frac{e(r)}{f(r)} \cos^2 \theta.
\end{align}
where $\mathcal{K}$ is the separation constant. Using the definition of the canonically conjugate momentum $p_\mu=g_{\mu \nu} \frac{dx^\nu}{d\tau}$, we derive the complete set of equations describing the photon motion 
\begin{align}
\label{dtdtau}
r^2\frac{dt}{d \tau}&=\frac{E r^2}{f(r)} -a L \frac{ e(r)}{ f(r)}\sin^2\theta, \\
\label{drdtau}
r^2\frac{dr}{d \tau}&=\sqrt{R(r)}, \\
\label{dthdtau}
r^2 \frac{d \theta}{d \tau}&=\sqrt{\Theta(\theta)}, \\
\label{dphidtau}
r^2\frac{d \phi}{d \tau}&=L+a E \frac{ e(r)}{f(r)}\sin^2\theta.
\end{align}
where $R(r)$ and $\Theta(\theta)$ are expressed as 
\begin{align}
\label{Rr}
R(r)&= E^2 r^4 \frac{g(r)}{f(r)}+2aELr^2\frac{g(r)e(r)}{f(r)}-r^2g(r)(L^2+\mathcal{K}), \\
\Theta(\theta) &= \mathcal{K} \csc^2 \theta-2a E L \frac{e(r)}{f(r)} \cot^2 \theta.
\end{align}
To examin the geometrical shapes of the shadow,  a suitable way would be to consider the effective potential which has the following form
\begin{equation}
V_{eff}(r)=-\left( \frac{dr}{d \tau} \right)^2=-E^2\frac{g(r)}{f(r)}+\frac{2aEL}{r^2}\left(g(r) - \sqrt{\frac{g(r)}{f(r)}} \right) +\frac{g(r)}{r^2}\left( L^2+\mathcal{K} \right).
\end{equation}
It is worth noting that the obtained potential matches the slowly rotating black hole in the absence of halo dark matter when taking $g(r)=f(r)=1-\frac{2M}{r}$.
To examine the behaviors of the photon sphere associated with the effective potential maximum value, we illustrate such quantity as a function of $r$ for different values of the spin parameter $a$, the halo  total mass $M$ and the length scale $a_0$ in figure \eqref{V}.

From figure \eqref{V}, we observe that the halo mass $M$ decreases the effective potential significantly for small values of $a_0$. For higher $a_0$ values, we notice that the effective potential decreases much slowly when $M$ increases. The same impact is observed when analyzing the photon sphere radius $r_0$ corresponding to the effective potential maximum value. Such radius increases when the total mass $M$ and the length scale $a_0$ are increased. Regarding the spin parameter $a$, the photon sphere size increases while the effective potential decreases when higher values of such a spin are considered.
\subsection{Black hole shadows}
\begin{figure}[H]
\begin{tabular}{lr}
\includegraphics[scale=0.45]{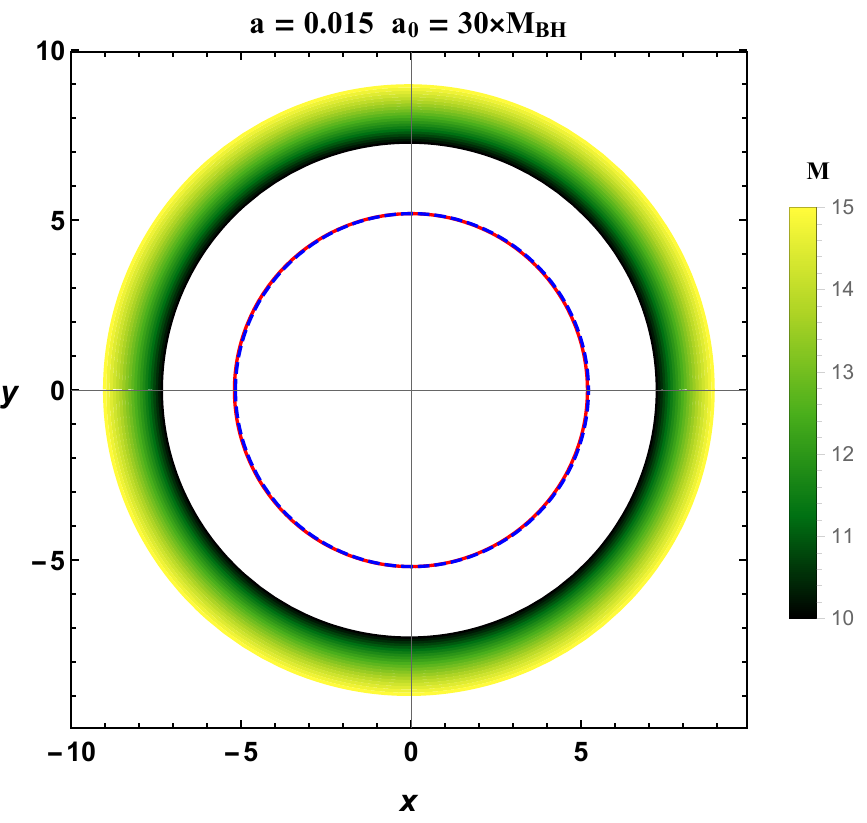}  & \hspace{1cm}\includegraphics[scale=0.45]{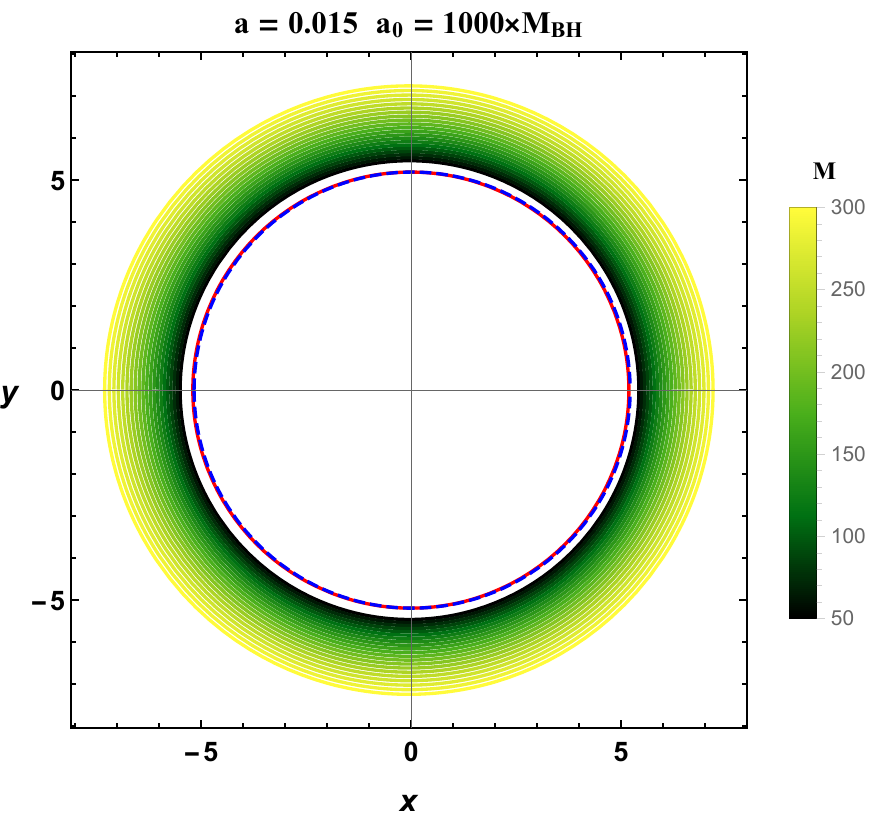}\\
\includegraphics[scale=0.45]{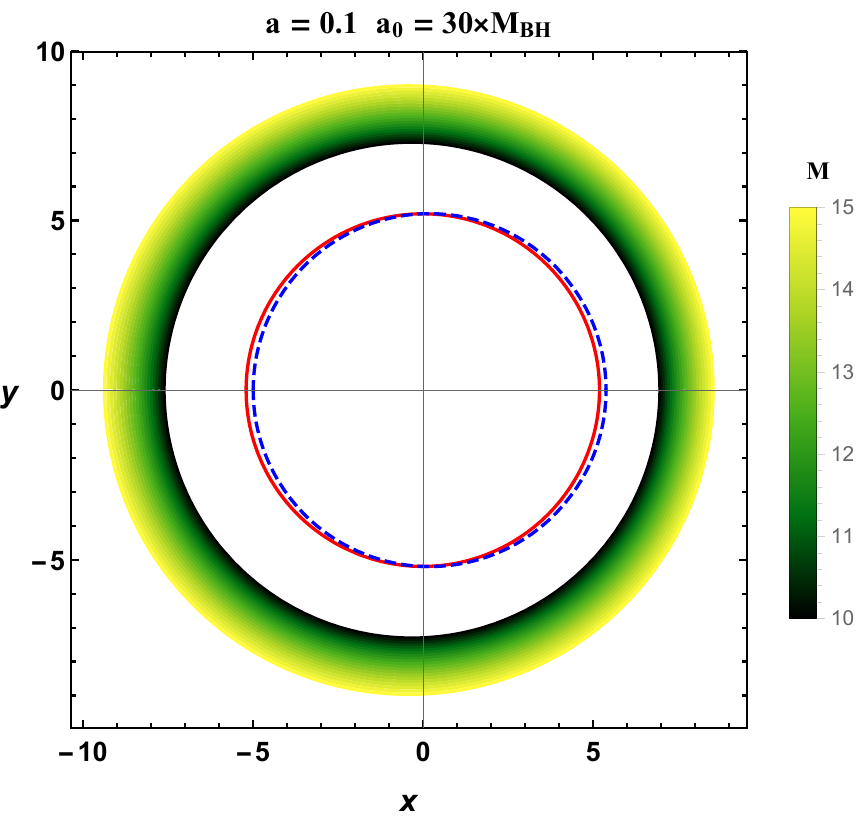} & \hspace{1cm} \includegraphics[scale=0.45]{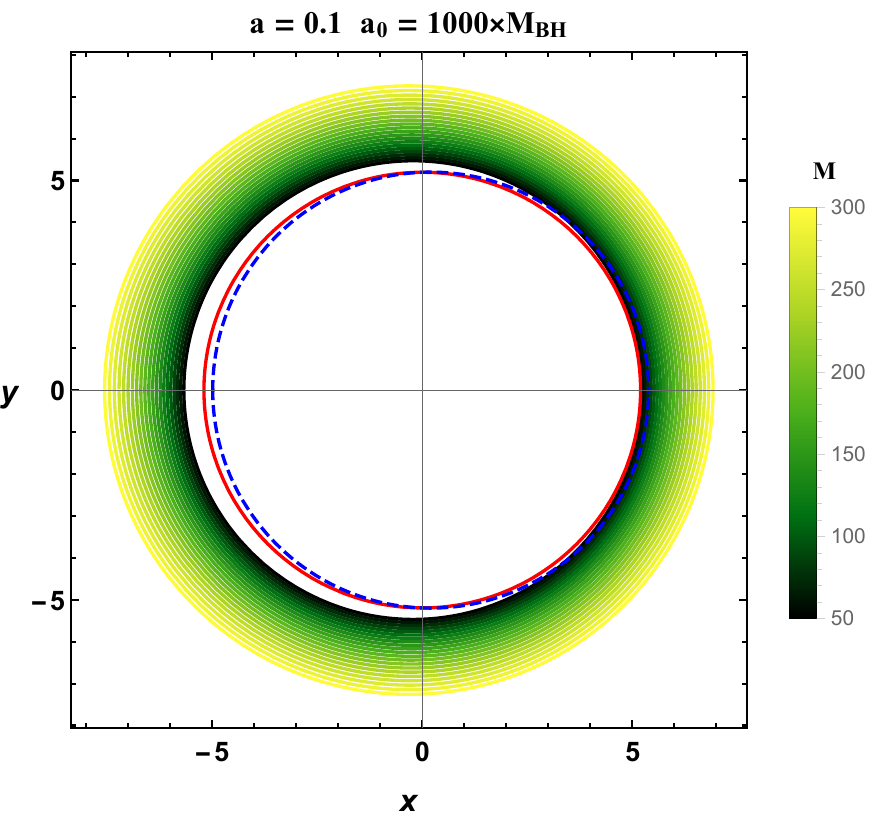}
\end{tabular}
\caption{\footnotesize Shadow for different values of the halo total mass $M$, length scale $a_0$,  the spin $a$ and fixed black hole mass $\left( M_{BH}=1 \right)$. The blue dashed curve correspond to the slowly-rotating Kerr black hole with a spin $a=0.015$ (top) and $a=0.1$ (bottom) while the red curve correspond to Schwarzschild case. }
\label{Shad}
\end{figure}

To explore the shadow geometrical shape in the slowly rotating regime, we introduce the following impact parameters
\begin{equation}
\xi=\frac{L}{E}, \quad \eta=\frac{\mathcal{K}}{E^2}.
\end{equation}
In this way, the function $R(r)$ given in \eqref{Rr} is rewritten as a function of these two impact parameters
\begin{equation}
R(r)=E^2 \left( r^4 \frac{g(r)}{f(r)}+2a \xi r^2\frac{g(r)e(r)}{f(r)}-r^2g(r)(\xi^2+\eta) \right).
\label{fRr}
\end{equation}
The critical unstable circular orbits can directly be derived from the following conditions
\begin{equation}
R(r) \big\vert_{r_0}=0, \quad \frac{d R(r)}{dr}\big\vert_{r_0}=0.
\label{Cond}
\end{equation}
With the use of \eqref{fRr} and \eqref{Cond}, we derive the impact parameters $\eta$ and $\xi$ that are expressed as
\begin{align}
\label{xi}
\xi = & \frac{r \left(2 f(r) -r f'(r)\right)}{2 a \left(e(r) f'(r)-f(r) e'(r)\right)}  ,\\
\label{eta}
\eta = &-\frac{r \left( r^3 f'(r)^2-4 r^2 f(r) f'(r)+4 r f(r)^2 \right)}{4 a^2 \left(e(r) f'(r)-f(r) e'(r)\right)^2}.
\end{align}
The allowed values of $\xi$ and $\eta$ rule the shadow geometrical shape. However, to picture the shadow as a distant observer sees it a better approach would be to consider the celestial coordinates $x$ and $y$ defined by
\begin{align}
x&= \lim_{r_* \to \infty} \left( -r^2_* \sin^2 \theta_0 \frac{d \phi}{dr} \right), \\
y&= \lim_{r_* \to \infty} r^2_* \frac{d \theta}{dr},
\end{align}
where $r_*$ is the distance between the black hole and the observer and $\theta_0$ is associated to the inclination angle between the line rotational axis of the black hole and the observer line of sight \cite{2S1}. As a function of the impact parameters, these two celestial coordinates can be written as
\begin{align}
\label{cel1}
x&= -\xi \csc \theta_0, \\
\label{cel2}
y&= \sqrt{\eta -\xi^2 \cot^2 \theta_0},
\end{align}
The shadow is then governed by the following equation
\begin{equation}
x^2+y^2=\xi^2+\eta,
\end{equation}
Considering that the observer is located at the equatorial plan $\left( \theta=\frac{\pi}{2} \right)$, we  illustrate the shadow in figure \eqref{Shad} for different values of the spin parameter $a$, the halo  total mass $M$ and the length scale $a_0$. \\
From figure \eqref{Shad}, we remark that the shadow size increases for higher values of the halo total mass $M$. However, when the length scale $a_0$ values are increased we observe that the shadow size is much smaller even if the halo mass takes significant values. For the particular value $a_0=1000$, the shadow radius is small and tends to the Schwarzschild case when $M$ take small values. An interesting result emerges for the slowly rotating black hole when it's surrounded by halo dark matter. Indeed, we notice by comparing the latter with the ordinary slowly rotating case (blue dashed circle), that the considered black hole is shifted from its center to the left side. Such behavior becomes obvious for higher values of the spin $a$.

\begin{figure}[H]
\begin{tabular}{lr}
\includegraphics[scale=0.5]{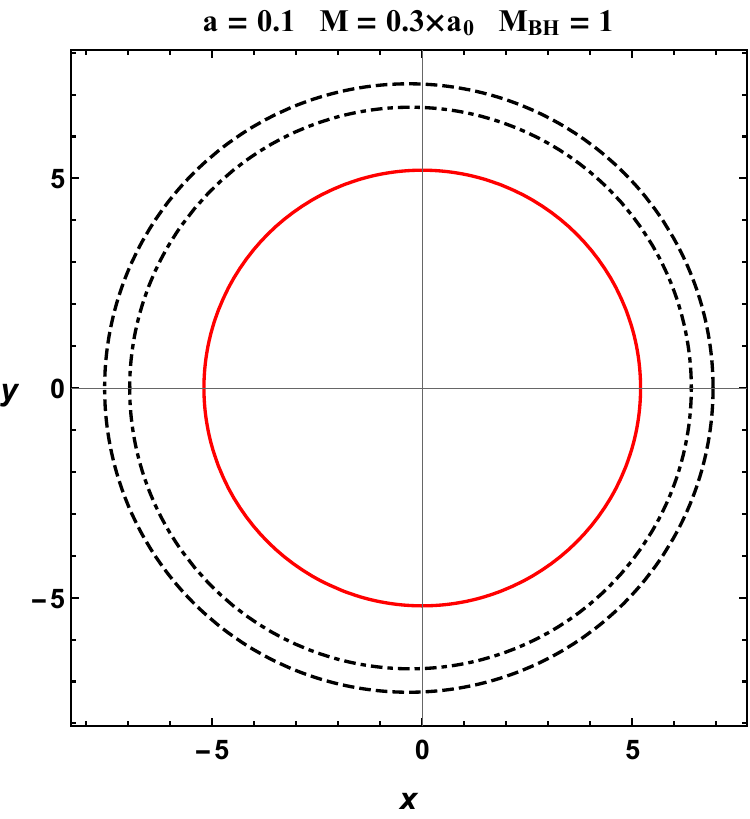} & \hspace{0.5cm} \includegraphics[scale=0.5]{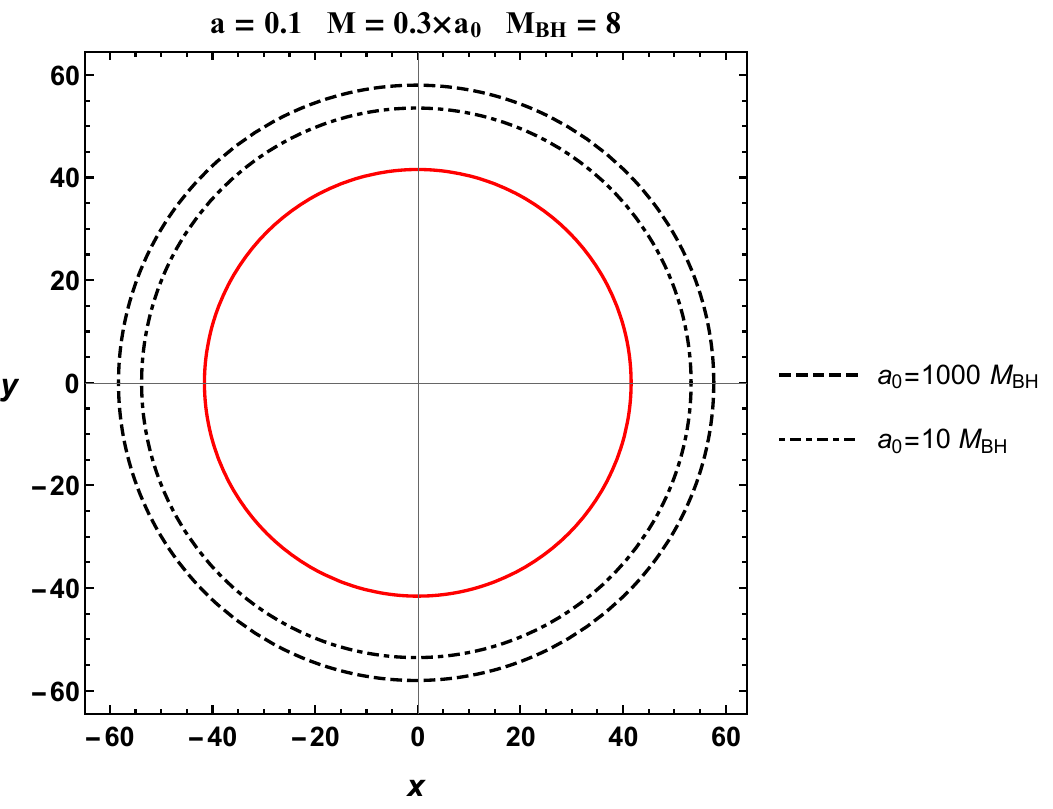}
\end{tabular}
\caption{ \footnotesize Shadow for different values of the black hole mass $M_{BH}$, halo total mass $M$, length scale $a_0$ and fixed spin $a$.}
\label{ShadM}
\end{figure}

Further investigation could be done to analyze the black hole shadow. In fact, it is interesting to check if the considered black hole can have the same shadow shape when the black hole mass $M_{BH}$ is varied. To do so, we plot the black hole shadow for different values of the black hole mass $M_{BH}$ in figure \eqref{ShadM}. In this figure, the spin has a fixed value $a=0.1$ while the parameters $a_0$ and $M$ take different values.

From such an illustration, we observe that when the black hole mass take important values, i.e $M_{BH}=8$, the shadow becomes circular even if the length scale value is increased. Thus, $a_0$ and $M$ control only the black hole size. However, for lower $M_{BH}$ values, we remark that the shadow is shifted from its center to the left side. This suggests that such behavior could be observed only for black holes with low mass values.
\subsection{Observational constraints}
\begin{figure}[H]
\begin{tabular}{lr }
\includegraphics[scale=0.4]{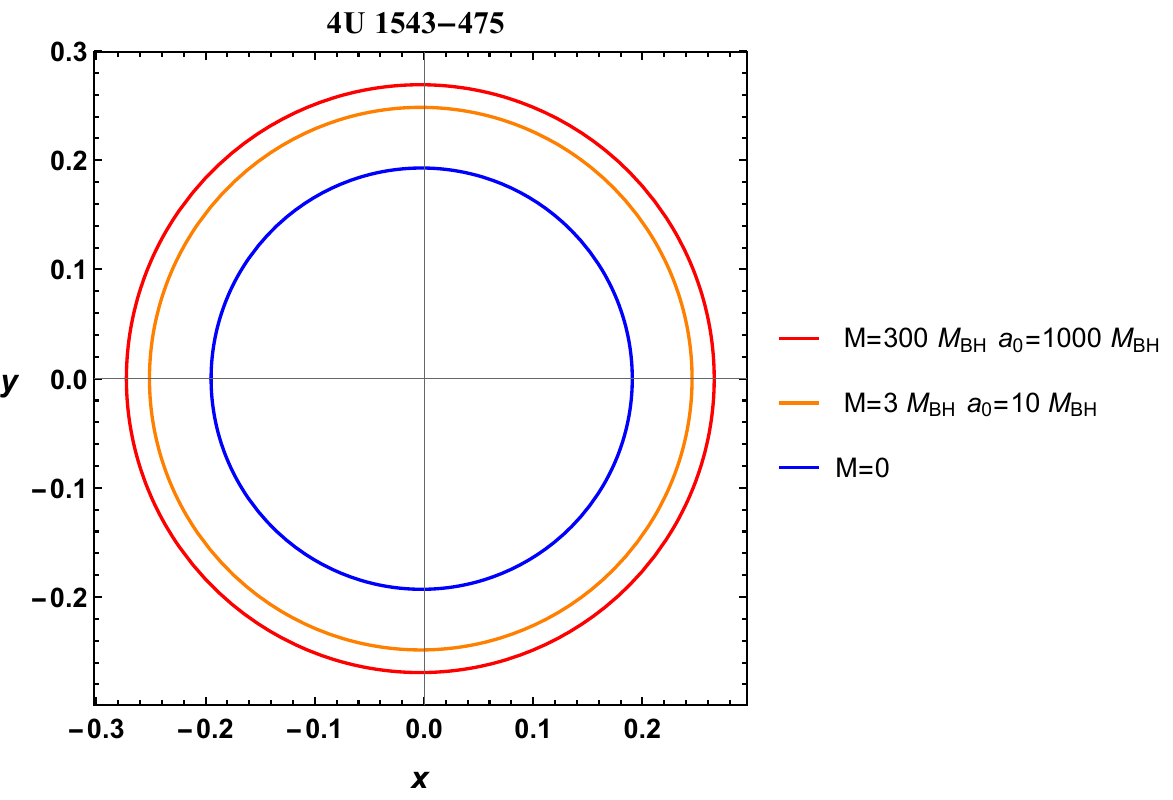} &\includegraphics[scale=0.4]{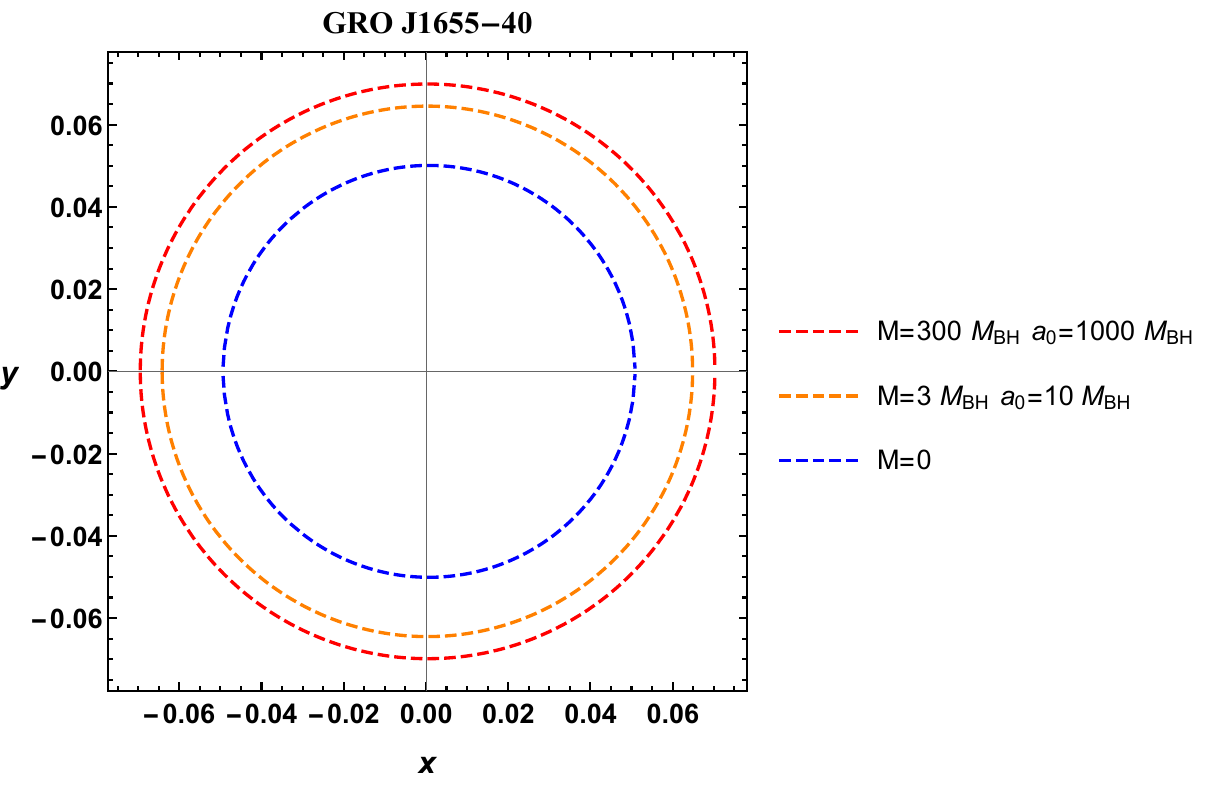}
\end{tabular}
\caption{ \footnotesize  (Left): 4U  1543-475 black hole shadow for different values of the halo mass and length scale. (Right): GRO J1655-40 black hole shadow for different values of the halo total mass and length scale. The blue shadow represents the slowly rotating black hole shadow when $M \to 0$. }
\label{EXP}
\end{figure}

To gain further insight of the halo total mass and the length scale effects on the black hole shadow, we consider two slowly rotating black holes in galactic nuclei. In our procedure, only slowly rotating black holes that satisfy $a^2 < a$ are taken into account. The first case correspond to 4U  1543-475 black hole with a spin $a=0.28$, a mass $M_{BH}= \left( 9.4 \pm 2.0 \right) M_{\odot}$ and an inclination angle $\theta_0= \left( 20.7^\circ \pm 1.0^\circ \right)$ while the second is associated to GRO J1655-40 black hole with a spin $a= \left(0.29 \pm 0.03 \right)$, a mass $M_{BH}= \left( 5.31 \pm 0.07 \right) M_{\odot}$ and an inclination angle $70^\circ <\theta_0 < 75^\circ$ \cite{GRO,4U}. 
To illustrate the shadow behavior, we rely on the equations \eqref{cel1} and \eqref{cel2} where the inclination angle is included. Such a behavior is illustrated in figure \eqref{EXP} for different values of the halo mass $M$ and length scale $a_0$.

As its expected, we observe from such a figure that the black hole shadow is circular due to the high values of the black hole mass. For both black hole, the shadow size increases when the halo total mass and length scale are increased.

\section{Deflection angle}
\label{v}
\begin{figure}[H]
\begin{tabular}{lr }
\includegraphics[scale=0.5]{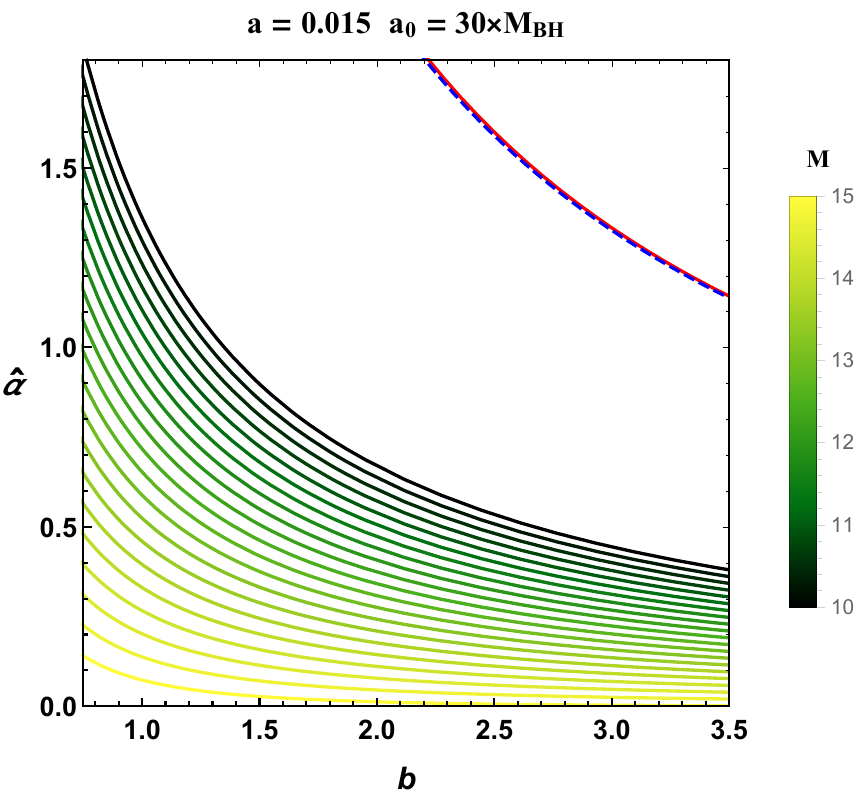} &\includegraphics[scale=0.5]{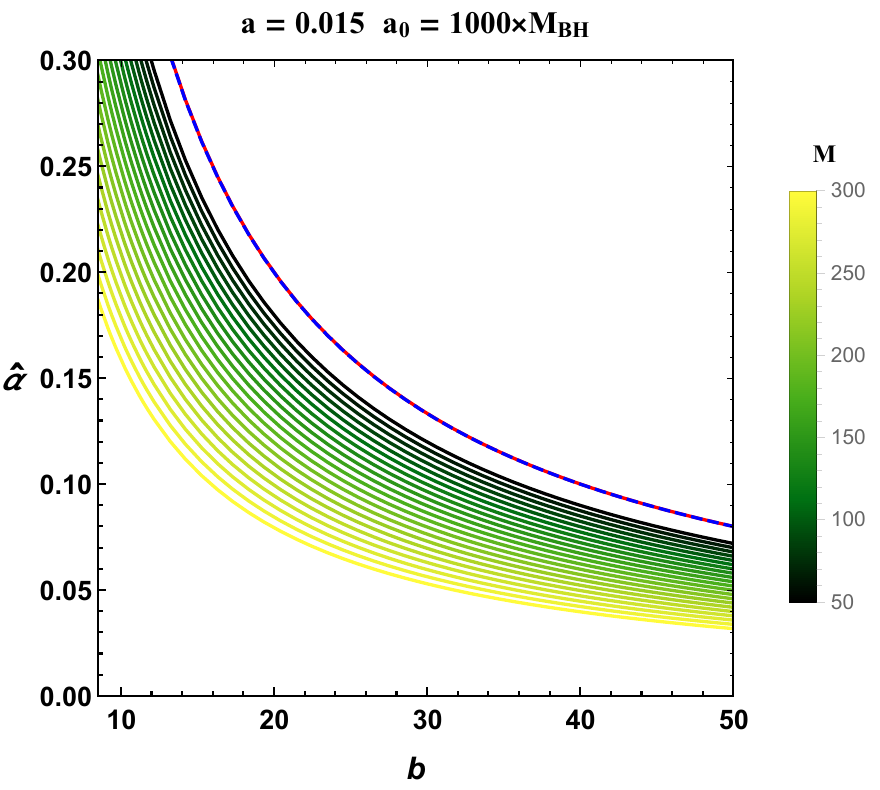}\\
\includegraphics[scale=0.5]{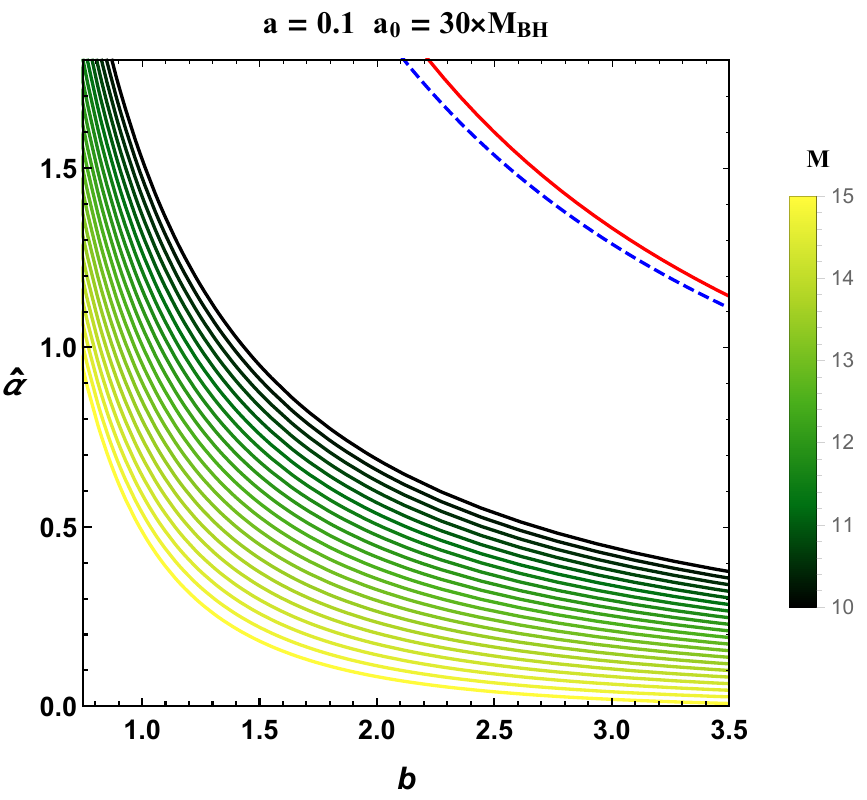} &\includegraphics[scale=0.5]{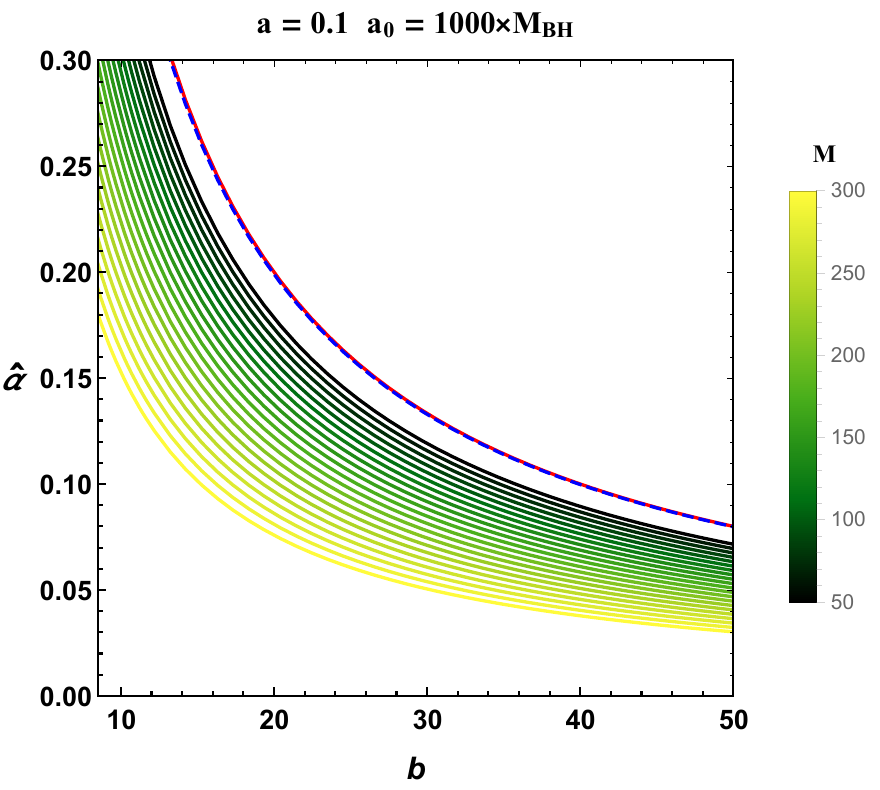}
\end{tabular}
\caption{ \footnotesize  Deflection angle for different values of the halo total mass $M$, length scale $a_0$,  the spin $a$ and fixed black hole mass $\left( M_{BH}=1 \right)$. The blue dashed curves is associated to the slowly-rotating Kerr black hole  with a spin $a=0.015$ (top) and $a=0.1$ (bottom) while the red curves represents Schwarzschild case. }
\label{Def}
\end{figure}

In this part of the paper, we explore the deflection angle of light by a slowly rotating black hole in galactic nuclei which is described by the metric \eqref{fmetric}. To obtain the needed results, we start by rewriting the metric in the equatorial plan $\left( \theta=\frac{\pi}{2} \right)$
\begin{equation}
ds^2=-\left[ f(r)+2ae(r) \frac{d\phi}{dt} \right] dt^2+\frac{dr^2}{g(r)}+r^2d\phi^2.
\end{equation}
In this equation $\frac{d\phi}{dt}$ can be calculated from equations \eqref{dtdtau} and \eqref{dphidtau}. Such a quantity can be expressed as a function of the impact parameter $b=\frac{E}{L}$ in the following way
\begin{equation}
\frac{d\phi}{dt}=\frac{f(r)+b a e(r)}{br^2-ae(r)}.
\end{equation}
Now, we define two new variables 
\begin{align}
dr_{*} &=\frac{dr}{\sqrt{g(r) \left( f(r)+2ae(r) \frac{d\phi}{dt} \right)}},\\
f(r_*) &=\frac{r}{\sqrt{f(r)+2ae(r) \frac{d\phi}{dt}}}.
\end{align}
In this way, one gets the optical metric for null geodesics $(ds^2=0)$
\begin{equation}
dt^2=g^{opt}_{mn}dx^m dx^n=dr_*^2+f(r_*)^2d\phi^2
\end{equation}
To obtain the deflection angle, we rely on the Gauss-Bonnet theorem that links the optical geometry to the topology. Such a theorem states 
\begin{equation}
\iint_{D_{R}} K dS+\oint_{\partial D_{R} }kdt+ \sum n_i=2\pi \chi \left( D_{R} \right).
\end{equation}
with $D_{R}$ being a non singular optical region, $\partial D_{R}$ its boundary, $k$ is the geodesic curvature and $K$ represents the Gaussian optical curvature. The geodesic curvature can be expressed as a function of a geodesic $\gamma_R$ as
\begin{equation}
k \left( \gamma_R \right)= \big\vert \nabla _{\gamma_R} \dot{\gamma_R} \big\vert.
\end{equation}
With the assumption that the geodesic $\gamma_R$ verifies $\gamma_R=R=cte$ the radial part of $k \left( \gamma_R \right)$ becomes
\begin{equation}
\left( \nabla _{\gamma_R}  \dot{\gamma_R} \right)^r= \dot{\gamma_R}^\phi + \partial_\phi \dot{\gamma_R}^r+ \Gamma^r_{\phi \phi} \left(\dot{\gamma_R}^\phi \right)^2.
\end{equation}
As it is shown in \cite{deflection}, the second term gives
\begin{equation}
\oint_{\partial D_{R} }kdt=\pi+\widehat{\alpha}.
\end{equation}
Besides, when the geometrical size $R$ of the optical region $D_R$ goes to infinity the jump angles $\alpha_S$ (source) and $\alpha_O$ (observer) are equal to $\frac{\pi}{2}$. The interior angles are $n_S=\pi-\alpha_S$ and $n_O=\pi-\alpha_O$. Thus, the deflection angle can be expressed rather simply when the linear approach of light ray is applied
\begin{equation}
\widehat{\alpha}=-\int_0^\pi \int_{\frac{b}{\sin \phi}}^\infty KdS,
\end{equation}
where $dS \simeq rdrd\phi$. In turn, the Gaussian optical curvature can be calculated with the relation 
\begin{equation}
K=\frac{\mathbf{R}}{2},
\end{equation}
which gives
\begin{equation}
K = -\frac{94 a b M M_{BH}}{a_0 r^5}+\frac{4 a b M}{a_0 r^4}+\frac{18 a b M_{BH}}{r^5}+\frac{4 M M_{BH}}{a_0 r^3}-\frac{2 M_{BH}}{r^3}+ \mathcal{O}\left(M_{BH}^2, \frac{1}{a_0^2}\right).
\end{equation}
The deflection angles is finally expressed as
\begin{equation}
\widehat{\alpha} = \frac{188 a M M_{BH}}{9 a_0 b^2}-\frac{\pi  a M}{a_0 b}-\frac{4 a M_{BH}}{b^2}-\frac{8 M M_{BH}}{a_0 b}+\frac{4 M_{BH}}{b} + \mathcal{O}\left(M_{BH}^2, \frac{1}{a_0^2}\right) ,
\label{DefA}
\end{equation}
where higher orders of $M_{BH}$ are omitted.  When $M \to 0$, the deflection angle \eqref{DefA} is given by
\begin{equation}
\widehat{\alpha} = -\frac{4 a M_{BH}}{b^2}+\frac{4 M_{BH}}{b},
\end{equation}
matching perfectly the rotating black hole deflection angle. To analyze the slowly rotating black hole deflection angle in galactic nuclei, we plot the associated behaviors as a function of the impact parameter $b$ in figure \eqref{Def} for different values of the halo total mass $M$, length scale $a_0$ and the spin $a$.

From these illustrations, we observe that the angle of deflection decreases when the halo total mass $M$ increases. For low values of the length scale $a_0$, the bending angle decreases for greater spin values. However, such an effect is not clearly seen for greater values of $a_0$. For the particular value $a_0=1000$, we remark that the deflection angle approach the Schwarzschild and slowly rotating black holes deflection angle for lower values of the halo mass $M$. Since, $M_{BH}<<M$ one should expect lower deflection angle values for black holes in galactic nuclei. We also notice a significant decrease of the deflection angle when photons approach the black hole.

\section{Conclusion and open questions}
\label{VI}

In this paper, we have constructed the slowly rotating case of an asymptotically flat supermassive black hole embedded in dark matter using Newman-Janis procedure. Such dark matter configurations, which is an extension of Einstein clusters with horizon, follows a Hernquist type density distribution observationally confirmed in elliptical galaxies. With the slowly rotating spacetime metric in hand, we have inspected the shadow and deflection angle behaviors of such black hole. By deriving the photon equation of motion, we have analyzed the circular orbits via the effective potential. Such an analysis has showed that the dark matter  length scale $a_0$ controls the effective potential values. Indeed, we have found that the halo total mass of the dark matter configuration decreases the effective potential significantly for low $a_0$ values while such an impact is diluted for high $a_0$ values. Concerning the shadow, we have showed that its size gets much smaller for high values of $a_0$. However, an opposite effect has been observed when the halo total mass $M$ is increased. Besides, the comparison of our case to the slowly rotating one have showed that the former exhibit a shadow shifted from its center to the left side due to the frame dragging effect. Then, we have investigated the mass variation influence on such shadow behaviors which gave a perfect circular shadow. Using such a result, we have provided a visualization of  $4U  1543-475$ and $GRO J1655-40$ black hole shadows. Finally, we have computed the deflection angle in the weak-limit approximation. From the illustrations of such quantity, we have showed that for high values of the halo total mass $M$ such quantity is decreased. Regarding the impact of the length scale $a_0$, we have obtained a decrease of the bending angle when the values of $a_0$ are low and the spin values are high. However, such an effect has not been clearly observed for greater $a_0$ values. We have also showed that one should expect low bending angle values for slowly rotating black holes in galactic nuclei.

With the use of the derived slowly rotating solution, further inspections could be considered. For instance, one can investigate the epicyclic oscillatory motion of test particles, quasinormal modes or thermodynamic aspects. We hope to address such situations in future works.
\appendix
\section{Energy-momentum, Einstein tensor and energy conditions}
\subsection{Energy-momentum and Einstein tensors}
\label{app1}
The energy-momentum tensor for a general spherically symmetric metric of the form
\begin{equation}
ds^2=-f(r) dt^2 + \frac{dr^2}{g(r)}+r^2 \left( d\theta^2 + \sin^2\theta  d\phi^2 \right),
\label{ffirst}
\end{equation}
can be given by 
\begin{equation}
T^{t}_{\, \, \, \, t}=-\rho, \quad T^{r}_{\, \, \, \, r}=p_r,\quad T^{\theta}_{\, \, \, \, \theta}= T^{\phi}_{\, \, \, \, \phi}=p_T.
\end{equation}
To determine the explicit expression of the energy-momentum tensor, we rewrite the metric functions $f(r)$ and $g(r)$ in the following way
\begin{equation}
g(r)=1-\frac{2m(r)}{r}, \quad f(r)=\frac{g(r)}{\left[j(r)\right]^2},
\end{equation}
with
\begin{equation}
j(r)=\sqrt{1-\frac{2 M r}{(a_0+r)^2}\left(1-\frac{2M_{BH}}{r}  \right)} \, e^{\frac{\pi}{2}\sqrt{\frac{M}{2a_0-M+4M_{BH}}}-\sqrt{\frac{M}{2a_0-M+4M_{BH}}}\arctan\left( \frac{r+a_0-M}{\sqrt{M\left( 2a_0-M+4M_{BH} \right)}} \right)}.
\end{equation}
Using the known definitions of $\rho, p_r$ and $p_T$, we obtain
\begin{align}
\label{frho}
\rho & =\frac{m'}{4 \pi r^2}=\frac{M \left(a_0+2M_{BH}\right)\left(1-\frac{2M_{BH}}{r} \right)  }{2\pi r\left(r+a_0\right)^3}, \\
p_r &=-\frac{m'}{4 \pi r^2}-\frac{(r-2m)j'}{4 \pi r^2 j}=0, \\
p_T &=-\frac{m''}{4 \pi r^2}+\frac{3r m'-r-m}{8 \pi r^2 j} j'+\frac{(r-2m)(j')^2}{4 \pi r j^2}-\frac{r-2m}{8 \pi r j}j'', \\
&=\frac{M(a_0+2M_{BH})(a_0^2 M_{BH}+2a_0 M_{BH} r+ M_{BH}r^2+M\left(r-2M_{BH} \right)^2 )}{4\pi r^2\left(a_0+r\right)^3 \left(a_0^2+4M M_{BH}+2a_0r-2Mr+r^2\right)},
\label{fp}
\end{align}
where $m'$ is associated to the first derivative of $m$ while $m''$ is the second derivative. Thus, the non-rotating energy-momentum tensor is 
\begin{equation}
T^{\mu}_{\, \, \, \, \nu}=diag\left(-\rho,0,p_T,p_T\right).
\label{TNR}
\end{equation}
The Newman-Janis algorithm application on the metric \eqref{ffirst} give rise to a slowly rotating black hole with a spacetime described by the following metric
\begin{equation}
ds^2=-f(r) dt^2 + \frac{dr^2}{g(r)}+h(r) d\Omega^2-2a \,  e(r) \sin^2 \theta dt d\phi.
\label{ffmetric}
\end{equation}
However, the energy-momentum tensor of such solution is different from the non-rotating one given by \eqref{TNR}. Using the function $j(r)$, we can rewrite the metric \eqref{ffmetric} as 
\begin{equation}
ds^2=-f(r) dt^2 + \frac{dr^2}{g(r)}+h(r) d\Omega^2-2a \left( \frac{1}{j(r)} -\frac{g(r)}{j(r)^2} \right) \sin^2 \theta dt d\phi.
\label{fmetricj}
\end{equation}
To establish the expression of the  energy-momentum tensor components, we introduce an orthonormal tetrad $e^\alpha_{\widehat{\alpha}}$ adapted to the metric \eqref{fmetricj}
\begin{equation}
e^\alpha_{\widehat{\alpha}}=\begin{pmatrix}
\frac{j}{\sqrt{1-\frac{2m}{r}}} & 0 & 0 & \frac{a \sin \theta}{r}  \\
0 & \sqrt{1-\frac{2m}{r}} & 0 & 0\\
0 & 0 & \frac{1}{r} & 0 \\
\frac{a}{r^2 \sqrt{1-\frac{2m}{r}}} & 0 & 0 & \frac{1}{r \sin \theta}
\end{pmatrix},
\end{equation}
such that $g_{\widehat{\alpha} \widehat{\beta}}=g_{\alpha \beta} \, e^\alpha_{\, \, \, \, \widehat{\alpha}} e^\beta_{ \, \, \, \,\widehat{\beta}}=diag(-1,1,1,1)$. In this way, the energy momentum tensor component forms become simple. However, the defined tetrad is not the principale frame of the energy momentum where it is diagonal. In this base, the latter is given by
\begin{equation}
T_{\widehat{\mu} \widehat{\nu}}=\begin{pmatrix}
-\widehat{u}_0 & 0 & 0 & \widehat{\sigma}_{30}  \\
0 & \widehat{u}_1 & \widehat{\sigma}_{12} & 0\\
0 & \widehat{\sigma}_{12} & \widehat{u}_2 & 0 \\
\widehat{\sigma}_{30} & 0 & 0 & \widehat{u}_3
\end{pmatrix},
\end{equation}
where $\widehat{\sigma}_{12}=\left( r\sqrt{1-\frac{2m}{r}} \right)  \sin \theta \, \,\widehat{u}_{12}, \widehat{\sigma}_{30}=\left( r\sqrt{1-\frac{2m}{r}} \right)\sin \theta \, \, \widehat{u}_{30} $ and the quantities $\widehat{u}_{i}, \widehat{u}_{ij}$ depend on $m, j$ and their derivatives $m', m'', j', j''$. According to \cite{A1art}, the final expressions of the  energy-momentum components can be obtained as a function of $\widehat{u}_{i}, \widehat{u}_{ij}$. In the case of a slowly rotating black hole, higher orders in $a$ should be omitted, yielding
\begin{align}
\widehat{u}_{0} & =-\frac{m'}{4 \pi r^2}=-\rho, \\
\widehat{u}_{1} & =-\frac{1}{4 \pi r^2 j}\left(j m'+j'(r-2m)  \right)=0, \\
\widehat{u}_{2} & =\frac{1}{8 \pi r^2 j^2}\left(-rj m''-rj(r-2m)j''+2r(r-2m)-j(m+r-3rm')j'  \right), \\
& =\frac{M(a_0+2M_{BH})(a_0^2 M_{BH}+2a_0 M_{BH} r+ M_{BH}r^2+M\left(r-2M_{BH} \right)^2 )}{4\pi r^2\left(a_0+r\right)^3 \left(a_0^2+4M M_{BH}+2a_0r-2Mr+r^2\right)}=p_T, \nonumber \\
\widehat{u}_{3} & =\frac{1}{8 \pi r^2 j^2}\left(r^2(2j'-j j'')+r(3j j' m'-j^2 m''-4 j'm +2 j j'' m-j j')-j j' m   \right), \\
& =\frac{M(a_0+2M_{BH})(a_0^2 M_{BH}+2a_0 M_{BH} r+ M_{BH}r^2+M\left(r-2M_{BH} \right)^2 )}{4\pi r^2\left(a_0+r\right)^3 \left(a_0^2+4M M_{BH}+2a_0r-2Mr+r^2\right)}=p_T, \nonumber \\
\widehat{u}_{30} &= \frac{a}{16 \pi r^4 j^2}\left(-2r j j'+r^2 j^2-r^2j j''-2j(j-1)  \right), \\
&=\frac{a}{16 \pi  r^4 (a_0+r)^2 \left(a_0^2+2 a_0 r+4 M M_{BH}-2 M r+r^2\right)^2} \nonumber \\
& \times \left\lbrace 4 M^2 r^2 (a_0+2 M_{BH})^2+M^2 r^2 (a_0+4M_{BH}-r)^2-2 M^2 r^2 (a_0+r) (a_0+4 M_{BH}-r)\right.  \nonumber \\
& -2 M r^2 (2 a_0+6  M_{BH}-r) \left(a_0^2+2 a_0 r+4 M  M_{BH}-2 M r+r^2\right)-2 M r^2 (a_0+r)^2 (a_0-M+r)  \nonumber \\
& -2 ( a_0+r)^2 \left( a_0^2+2  a_0 r+4 M  M_{BH}-2 M r+r^2\right)^2 -M^2 r^2 (a_0+r)^2  \nonumber \\
& +4 M r^3 (a_0+2 M_{BH}) \left(a_0^2+2 a_0 r+4 M M_{BH}-2 M r+r^2\right)^{3/2} \nonumber \\
& \times \exp \left(\frac{1}{2} \sqrt{\frac{M}{2 a_0-M+4 M_{BH}}} \left(\pi -2 \arctan\left(\frac{a_0-M+r}{\sqrt{M (2 a_0-M+4 M_{BH})}}\right)\right)\right) \nonumber \\
& +2 (a_0+r)^3 \left(a_0^2+2 a_0 r+4 M M_{BH}-2 M r+r^2\right)^{3/2} \nonumber \\
& \left. \times \exp \left(-\frac{1}{2} \sqrt{\frac{M}{2a_0-M+4 M_{BH}}} \left(\pi -2 \arctan\left(\frac{a_0-M+r}{\sqrt{M (2a_0-M+4 M_{BH})}}\right)\right)\right) \right\rbrace, \nonumber \\
\widehat{u}_{12} & =0.
\end{align}
Regarding the components of the energy-momentum tensor, we obtain
\begin{align}
T^t_{\, \, \, \, t}& =\frac{1}{r^2 j} \left( a \sin^2 \theta \left(r^2(1+j)-2rm \right)\widehat{u}_{30}+r^4 j^2 \widehat{u}_{0} \right)\simeq \widehat{u}_{0}=-\rho, \\
T^t_{\, \, \, \, \phi}& =\frac{\sin^2 \theta}{r^2 j} \left( a r^2 j \left(\widehat{u}_{3}-\widehat{u}_{0} \right)-r^4 j^2 \widehat{u}_{30} \right), \\
&= \frac{a \sin ^2(\theta )}{16 \pi  r^2 (a_0+r)^3 \left(a_0^2+2 a_0 r+4 M M_{BH}-2 M r+r^2\right)^2} \times \nonumber \\
& \left\lbrace \exp \left(\alpha \right)\sqrt{(a_0+r)^2+M (4 M_{BH}-2 r)}  \right.  \nonumber \\
& \times \left[ -4 M^2 r^2 (a_0+2 M_{BH})^2-M^2 r^2 (a_0+4 M_{BH}-r)^2+2 M^2 r^2 (a_0+r) (a_0+4 M_{BH}-r)\right. \nonumber \\
& +2 M r^2 (2 a_0+6 M_{BH}-r) \left(a_0^2+2 a_0 r+4 M M_{BH}-2 M r+r^2\right)+2 M r^2 (a_0+r)^2 (a_0-M+r) \nonumber \\
& +M^2 r^2 (a_0+r)^2 +2 (a_0+r)^2 \left(a_0^2+2 a_0 r+4 M M_{BH}-2 M r+r^2\right)^2 \nonumber \\
& -2 (a_0+r)^3 \left(a_0^2+2 a_0 r+4 M M_{BH}-2 M r+r^2\right)^{3/2} \times  \exp \left(-\alpha \right) \nonumber  \\
& \left. -4 M r^3 (a_0+2M_{BH}) \left(a_0^2+2 a_0 r+4 M M_{BH}-2 M r+r^2\right)^{3/2} \times \exp \left(\alpha \right) \right] \nonumber \\
& +4 M (a_0+2 M_{BH}) \left(a_0^2+2 a_0 r+4 M M_{BH}-2 M r+r^2\right) \nonumber \\
& \times \left. \left(a_0^2 (2 r-3 M_{BH})+a_0 \left(4 r^2-6 M_{BH} r\right)-3 M (r-2 M_{BH})^2+r^2 (2 r-3 M_{BH})\right) \right\rbrace, \nonumber \\
T^\phi_{\, \, \, \, t}& = \frac{1}{r^2 j}\left( r ( r-2m) \widehat{u}_{30}-a(\widehat{u}_3 - \widehat{u}_0)\right), \\
&= \frac{a \exp \left(-\alpha \right)}{16 \pi  r^5 (a_0+r)^3 \left(a_0^2+2a_0 r+4 M M_{BH}-2 M r+r^2\right) \sqrt{(a_0+r)^2+M (4M_{BH}-2 r)}} \times \nonumber \\
&  \left\lbrace 4 M r (a_0+2 M_{BH}) (a_0+r) \left[ a_0^2 (3 M_{BH}-2 r)+2 a_0 r (3 M_{BH}-2 r) \right.   \right. \nonumber \\
& \qquad \qquad \qquad \qquad \qquad \qquad \qquad \qquad \qquad \left. +3 M (r-2 M_{BH})^2+r^2 (3 M_{BH}-2 r)\right] \nonumber \\
& +(r-2 M_{BH})\left[4 M^2 r^2 (a_0+2 M_{BH})^2+M^2 r^2 (a_0+4 M_{BH}-r)^2-2 M^2 r^2 (a_0+r) (a_0+4 M_{BH}-r)  \right. \nonumber \\
& -2 M r^2 (2 a_0+6  M_{BH}-r) \left(a_0^2+2 a_0 r+4 M  M_{BH}-2 M r+r^2\right)-M^2 r^2 (a_0+r)^2 \nonumber \\
& -2 M r^2 (a_0+r)^2 (a_0-M+r) -2 (a_0+r)^2 \left(a_0^2+2 a_0 r+4 M M_{BH}-2 M r+r^2\right)^2 \nonumber \\
& +4 M r^3 (a_0+2 M_{BH}) \left(a_0^2+2 a_0 r+4 M M_{BH}-2 M r+r^2\right)^{3/2} \times  \exp \left(\alpha \right) \nonumber \\
& \left. \left.  +2 (a_0+r)^3 \left(a_0^2+2 a_0 r+4 M M_{BH}-2 M r+r^2\right)^{3/2} \times  \exp \left(-\alpha\right)    \right] \right\rbrace, \nonumber \\
T^\phi_{\, \, \, \, \phi}&=\frac{1}{r^2 j} \left(-a r \sin^2 \theta \left(r(1+j)-2m \right) \widehat{u}_{30}+r^2 j \widehat{u}_3 \right) \simeq \widehat{u}_3=p_T,  \quad T^r_{\, \, \, \, r} = \widehat{u}_1=0, \\
T^r_{\, \, \, \, \theta}&= \left( r^2-2r m \right) \widehat{u}_{12}\sin \theta=0, \quad T^\theta_{\, \, \, \, r} = \widehat{u}_{12}\sin \theta=0, \quad T^\theta_{\, \, \, \, \theta}= \widehat{u}_{2} = p_T,
\end{align}
where $\alpha = \frac{1}{2} \sqrt{\frac{M}{2 a_0-M+4M_{BH}}} \left(\pi -2 \arctan\left(\frac{a_0-M+r}{\sqrt{M (2 a_0-M+4 M_{BH})}}\right)\right)$. We omitted the terms with $a \times \widehat{u}_{30}$ since they are proportional to $a^2$. If we set $T^t_{\, \, \, \, \phi} = a \chi \left(r,\theta \right) $ and $T^\phi_{\, \, \, \, t}= a\Psi(r)  $, we can write the energy momentum tensor as follows
\begin{equation}
T^\mu_\nu=\begin{bmatrix}
-\rho & 0 & 0 & a  \chi \left(r,\theta \right) \\
0 & 0 & 0 & 0 \\
0 & 0 & p_T & 0 \\
a \Psi(r) & 0 & 0 & p_T 
\end{bmatrix}.
\end{equation}
Thus, such a form can describe a slowly rotating black hole in active galactic nuclei. Taking the limit $a \to 0$,  we recover the non rotating energy momentum tensor given in Eq.\eqref{TNR}. It is worth noting that the Einstein tensor can also be calculated in the slowly rotating regime. Indeed, with the use of the metric provided in equation \eqref{fmetricj}, we obtain the following Einstein tensor components
\begin{align}
G_{t}^{t} &=\frac{r g^{\prime}(r)+g(r)-1}{r^2}, \nonumber \\
G_{t}^{\phi }& =\frac{a }{4 r^4 f(r)^2}  \left(2 r^2 f(r)^2 g(r) e^{\prime \prime}(r)-r^2 f(r) g(r) e^{\prime}(r) f^{\prime}(r)+r^2 f(r)^2 e^{\prime}(r) g^{\prime}(r) \right. \nonumber \\
& \left. -2 r^2 e(r) f(r) g(r) f^{\prime \prime}(r) -r^2 e(r) f(r) f^{\prime}(r) g^{\prime}(r)+r^2 e(r) g(r) f^{\prime}(r)^2-4 e(r) f(r)^2\right), \nonumber \\
G_{\phi }^{t} & =-\frac{a \sin ^2(\theta )}{4 r^2 f(r)^2} \left(2 r^2 f(r) g(r) e^{\prime \prime}(r)-r^2 g(r) e^{\prime}(r) f^{\prime}(r)+r^2 f(r) e^{\prime}(r) g^{\prime}(r) \right.  \nonumber \\
&\left. +2 r e(r) g(r) f^{\prime}(r)-2 r e(r) f(r) g^{\prime}(r)-4 e(r) f(r) g(r)\right),  \nonumber \\
G_{\theta }^{\theta } &=\frac{2 r f(r) g(r) f^{\prime \prime}(r)+r f(r) f^{\prime}(r) g^{\prime}(r)+2 f(r) g(r) f^{\prime}(r)-r g(r) f^{\prime}(r)^2+2 f(r)^2 g^{\prime}(r)}{4 r f(r)^2},  \nonumber \\
G_{\phi }^{\phi } &=\frac{2 r f(r) g(r) f^{\prime \prime}(r)+r f(r) f^{\prime}(r) g^{\prime}(r)+2 f(r) g(r) f^{\prime}(r)-r g(r) f^{\prime}(r)^2+2 f(r)^2 g^{\prime}(r)}{4 r f(r)^2}.
\end{align}
From these expressions, we remark that $G_{\phi }^{t}$ depend on $a, \theta$ and $r$ and that $G_{t}^{\phi }$ depend on $a$ and $r$ which agree with the expression of the  energy-momentum tensor given above.  By taking the limit $a \to 0$, $G_{\phi }^{t}$ and $G_{t}^{\phi }$ go to zero and the non rotating, symmetrical and diagonal Einstein tensor can be recovered. Besides, by computing the expression of $G_{t}^{t}$ and $G_{\theta }^{\theta }$ or $G_{\phi }^{\phi }$, we obtain   
\begin{align}
G_{t}^{t} &= \frac{4 M (a_0+2 M_{BH}) (2 M_{BH}-r)}{r^2 (a_0+r)^3}, \nonumber \\
G_{\phi }^{\phi }=G_{\theta }^{\theta } &= \frac{2 M (a_0+2 M_{BH}) \left(a_0^2 M_{BH}+2 a_0 M_{BH} r+M (r-2 M_{BH})^2+M_{BH} r^2\right)}{r^2 (a_0+r)^3 \left(a_0^2+2 a_0 r+4 M M_{BH}-2 M r+r^2\right)}.
\end{align}
Since the Einstein equation is given by $T^\mu_\nu=\frac{1}{8 \pi} G^\mu_\nu$, we  conclude that the computations of $G^\mu_\nu$ agree with the expressions of the quantities $\rho$ in \eqref{frho} and $p_T$ given in equation \eqref{fp} and also the ones derived through the calculation of the energy-momentum tensor. 

\subsection{Weak Energy Condition}
\begin{figure}[h]
 \begin{tabular}{lr }
 \includegraphics[scale=0.45]{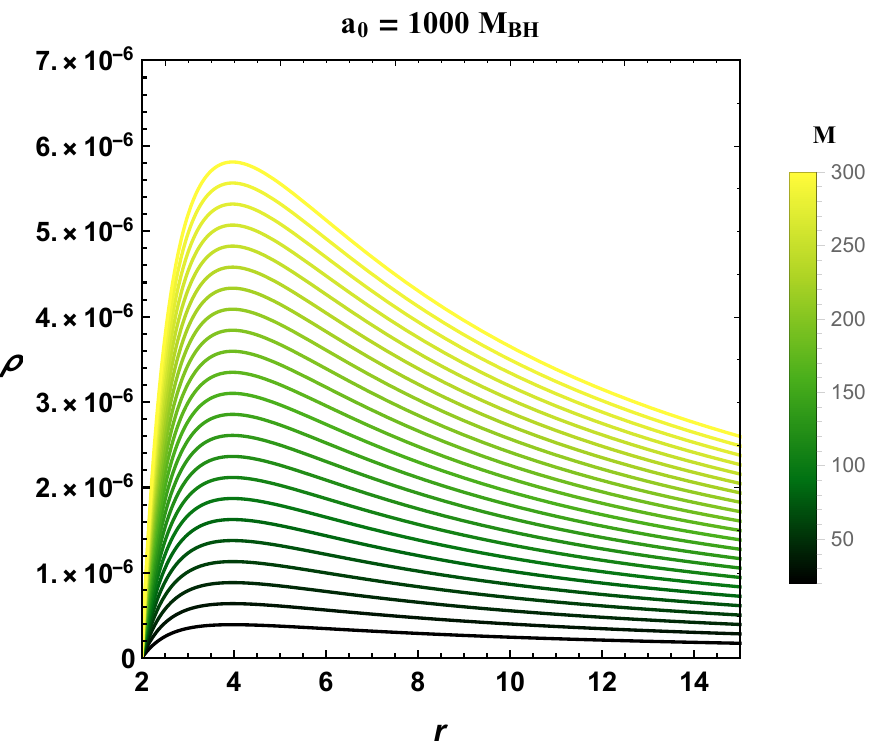} &\includegraphics[scale=0.45]{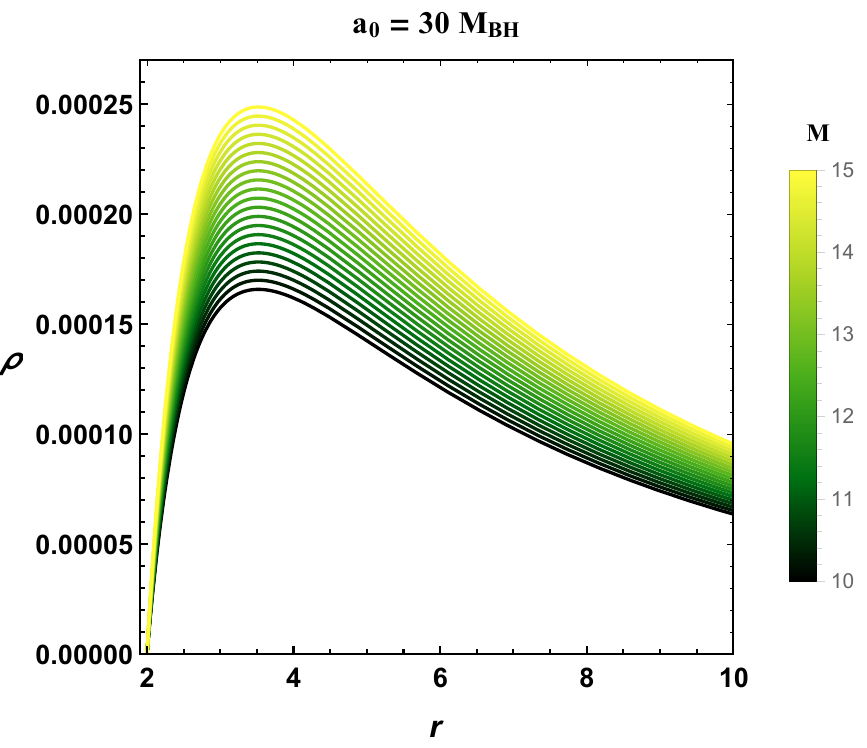}\\
 \includegraphics[scale=0.5]{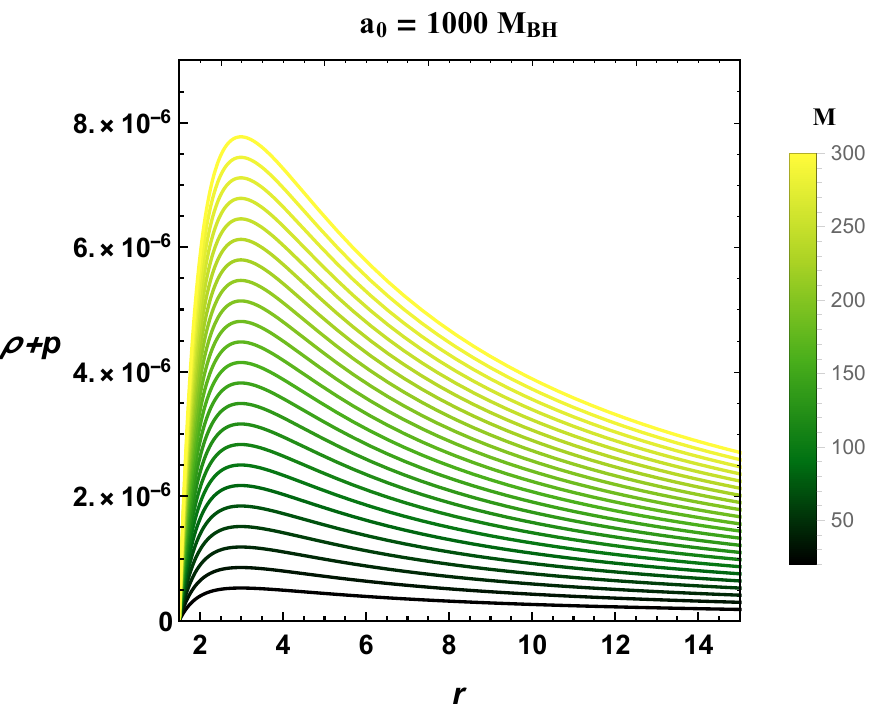} &\includegraphics[scale=0.5]{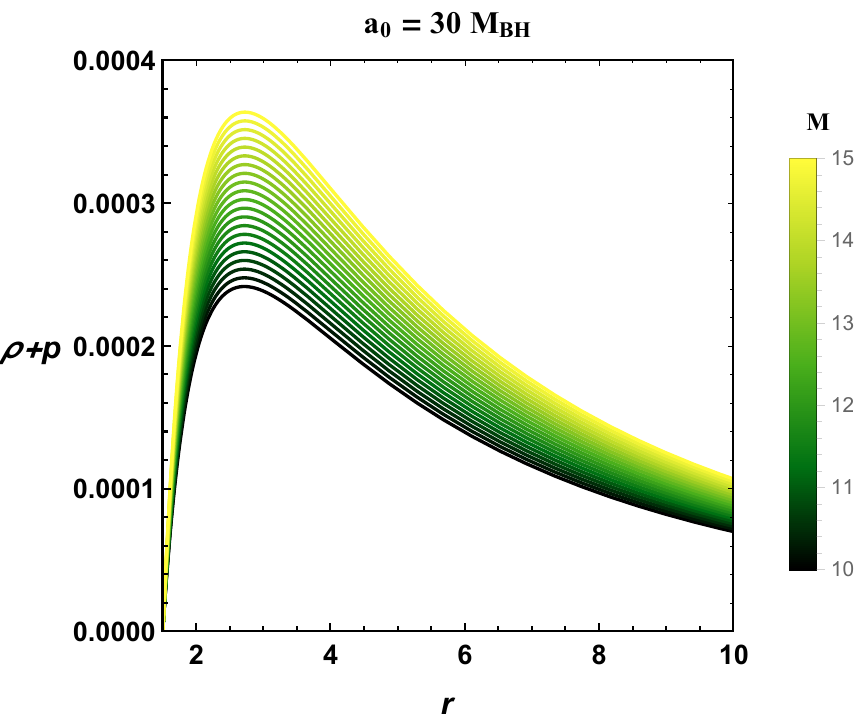}
 \end{tabular}
 \caption{ \footnotesize Dependence of matter density $\rho$ and $\rho+p_T$ on the radius for the slowly rotating black hole.  }
 \label{weak}
 \end{figure}

The Weak Energy Condition can be used in singularity theorems associated with black holes. In specific terms, the later is exploited to show that under certain conditions, the formation of singularities is indispensable.  In the present work, we consider the last  energy momentum tensor $T^\mu_\nu$ associated with  the slowly rotating solution. Using this tensor, we have derived the quantities $\rho$ and $p_T$ which are given by
\begin{align}
\label{rf}
\rho & =\frac{M \left(a_0+2M_{BH}\right)\left(1-\frac{2M_{BH}}{r} \right)  }{2\pi r\left(r+a_0\right)^3}, \\
p_T &=\frac{M(a_0+2M_{BH})(a_0^2 M_{BH}+2a_0 M_{BH} r+ M_{BH}r^2+M\left(r-2M_{BH} \right)^2 )}{4\pi r^2\left(a_0+r\right)^3 \left(a_0^2+4M M_{BH}+2a_0r-2Mr+r^2\right)}.
\label{ptf}
\end{align}
According to the weak energy condition, all classical matter must be non-negative when by any observer in space-time [48], i.e
\begin{equation*}
T_{\mu \nu} \xi^\mu \xi^\nu \ge 0
\end{equation*}
for all the time-like vectors $\xi^\mu$. By decomposing the energy-momentum tensor, we find that the weak energy condition can be written as 
 \begin{equation*}
\rho \ge 0, \rho + p_T \ge 0.
\end{equation*}
To verify the weak energy conditions, we illustrate the variation of $\rho$ and  $\rho + p_T$  as function of the radial coordinate $r$ according to equations \eqref{rf} and \eqref{ptf}. Indeed,  in the  figure \eqref{weak} we present the variation of these quantities  for different values of the parameters $a_0$ and $M$.  It can be shown that the weak energy condition is violated near the origin which is a common feature for all the rotating black holes. Further investigation shows that the quantity $\rho$ is positive when
\begin{equation}
r \ge 2M_{BH},
\end{equation}
while $\rho + p_T$ is positive when
\begin{equation}
r \ge \frac{1}{6} \left(\frac{4 a_{0}^2+12 a_{0} (M_{BH}-2 M)+9 \left(M^2-6 M M_{BH}+M_{BH}^2\right)}{\sqrt[3]{A}}+\sqrt[3]{A}-4 a_{0}+3 M+3 M_{BH}\right),
\end{equation}
where we have
\begin{align}
A & =  9 M \left(10 a_{0}^2+42 a_{0} M_{BH}+45 M_{BH}^2\right)-27 M^2 (4 a_{0}+9 M_{BH})+(2 a_{0}+3 M_{BH})^3+27 M^3 \nonumber \\
& + 18 \sqrt{M} (a_{0}+2 M_{BH})  \left( 9 M_{BH} \left(4 a_{0}^2+2 a_{0} M-M^2\right)+a_{0}^2 (8 a_{0}-3 M) \right.  \nonumber \\
& \left.  + 54 M_{BH}^2 (a_{0}+M)+27 M_{BH}^3 \right)^{1/2}. 
\end{align}
Analysing  $\rho + p_T$  numerically for the different values of the involved parameters $a_0$ and $M$, we find that the quantity $\rho + p_T$ is positive for the values $r > 1.5 $ which is observable from figure \eqref{weak}. Finally, we remark that $\rho \underset{r \to+\infty}{\longrightarrow} 0$ and  $\rho + p_T \underset{r \to+\infty}{\longrightarrow} 0$.

\newpage
\section*{Conflicts of interest}
The authors declare that they have no known competing interests or personal relationships that could have appeared to influence the work reported in this paper.

\section*{Data Availability}
No datasets were generated or analysed during the current study.


\begin{thebibliography}{10}
\bibitem{O0} J. B. Hartle, T. Dray, \textit{Gravity: An introduction to Einstein's general relativity}, Amer.
J. Phys.\textbf{ 71}, (2003), 1086-1087. \\
S. W. Hawking, W. Israel, \textit{General Relativity: an Einstein Centenary Survey}, UK Cam-
bridge University Press (2010).

\bibitem{O1} J. M. Bardeen, B. Carter, S. W. Hawking, \textit{The four laws of black hole mechanics},
Commun. Math. Phys. \textbf{31} (1973), 161.

\bibitem{O2} J. J. Fern\'andez-Melgarejo,  E. Torrente-Lujan, \textit{$\mathcal{N}=2$ SUGRA BPS multi-center solutions, quadratic prepotentials and Freudenthal transformations}, J. High Ener. Phys. \textbf{5} (2014), 81.

\bibitem{O3} J. M. Maldacena, \textit{Black holes in string theory}, Ph. D. Thesis, Princeton University (1996), hep-th/9607235.

\bibitem{O4}  L. Borsten, M. J. Duff, J. J. Fern\'andez-Melgarejo,  A. Marrani,  E. Torrente-Lujan, \textit{Black holes and general Freudenthal transformations}, J. High Ener. Phys. \textbf{1907} (2019), 070.

\bibitem{O5} O. Aharony,  S. S. Gubser,  J. Maldacena,  H. Ooguri,  Y.  Oz, \textit{Large N field theories, string theory and gravity}, Phys. Repo. \textbf{323}(3-4) (2000), 183-386.

\bibitem{O6} A. Belhaj, M. Chabab, H. El Moumni, M. B. Sedra, \textit{On thermodynamics of AdS black holes in arbitrary dimensions}, Chin. Phys. Lett. \textbf{29} (2012), 100401.
\bibitem{O6-1} A. Belhaj, A. El Balali, W. El Hadri, E.  Torrente-Lujan,  \textit{On universal constants of AdS black holes from Hawking-Page phase transition}, Phys. Lett. B \textbf{811} (2020), 135871. 
\bibitem{O6-2}  A. Belhaj, A. El Balali, W. El Hadri, H. El Moumni, M. A. Essebani, M. B. Sedra, \textit{On Phase Transition Behaviors of Kerr-Sen Black Hole}, Inter. Jour. Geo. Meth. Mod. Phys. \textbf{17} (2020), 2050169.
\bibitem{O6-3}  A. Belhaj, M. Chabab, H. El Moumni, K. Masmar, M. B. Sedra, A. Segui, \textit{On heat properties of AdS black holes in higher dimensions}, J. High Ener. Phys. \textbf{05} (2015), 149. 
\bibitem{O6-4} S. W. Hawking, D. N. Page, \textit{Thermodynamics of black holes in anti-de Sitter space},  Commun.\ Math.\ Phys.\ \textbf{87}(4) (1983), 577-588.
\bibitem{O7}  E. Torrente-Lujan, \textit{Smarr mass formulas for BPS multicenter black holes}, Phys. Lett. B\textbf{798} (2019), 135019.

\bibitem{O8} J. M. Maldacena, \textit{The large-N limit of superconformal field theories and supergravity}, Inter. J. Theor. Phys. \textbf{38}(4) (1999), 1113.

\bibitem{O9} E. Witten, \textit{Anti de Sitter space and holography}, Adv. Theor. Math. Phys. \textbf{2} (1998), 253-291.

\bibitem{O10} A. Belhaj,  A. El Balali,  W. El Hadri,  H. El Moumni,  M. B.  Sedra, \textit{Dark energy effects on charged and rotating black holes}, Eur. Phys. J. Plus  \textbf{134}(9) (2019), 422.

\bibitem{I1} K. Akiyama, et al., \textit{Event Horizon Telescope Collaboration}, Astrophys. J. \textbf{875} (1) (2019), p. L1.
\bibitem{I2}  K. Akiyama, et al., \textit{The Event Horizon Telescope Collaboration}, Astrophys. J. Lett.\textbf{910} (2021), p. L12. 

\bibitem{I3} K. Akiyama, et al., \textit{The Event Horizon Telescope Collaboration}, Astrophys. J. Lett. \textbf{910} (2021), p. L13.


\bibitem{I4} J. L. Synge,   \textit{The escape of photons from gravitationally intense stars}, Mont. Not. Roy. Astro. Soc. \textbf{131} (1966), 463-466.

\bibitem{I5} J. M. Bardeen, \textit{Les Houches Summer School of Theoretical Physics: Black Holes}, (New York: Gordon and Breach, Science Publishers, Inc. (1973), p. 219. 

\bibitem{I6} S. W. Wei, Y. C. Zou, Y. X. Liu, R. B Mann, \textit{Curvature radius and Kerr black hole shadow}, J. Cosmol. Astropart. Phys. \textbf{2019} (08), 30.

\bibitem{I7} M. Ghasemi-Nodehi, M. Azreg-Aınou, K. Jusufi, M. Jamil, \textit{Shadow, quasinormal modes, and quasiperiodic oscillations of rotating Kaluza-Klein black holes}, Phys. Rev. D \textbf{102} (2020), 104032.

\bibitem{I8} P. Bambhaniya, D. Dey, A. B. Joshi, P. S. Joshi, D. N. Solanki, A. Mehta, \textit{Shadows and negative precession in non-Kerr spacetime}, Phys. Rev. D \textbf{103} (2021), 084005.

\bibitem{I9} M. Fathi, M. Olivares, J. R. Villanueva, \textit{Ergosphere, photon region structure, and the shadow of a rotating charged Weyl black hole}, Galaxies \textbf{9} (2021), 43.

\bibitem{I10} F. Atamurotov, S. G. Ghosh, B. Ahmedov, \textit{Horizon structure of rotating Einstein–Born–Infeld black holes and shadow}, Eur. Phys. J. C \textbf{76} (2016), 1.

\bibitem{I11} F. Atamurotov, B. Ahmedov, A. Abdujabbarov, \textit{Optical properties of black holes in the presence of a plasma: The shadow}, Phys. Rev. D \textbf{92} (2015), 084005.

\bibitem{I12} U. Papnoi, F. Atamurotov, S. G. Ghosh, B. Ahmedov, \textit{Shadow of five-dimensional rotating Myers-Perry black hole},  Phys. Rev. D \textbf{90} (2014), 024073.

\bibitem{I13} F. Atamurotov, A. Abdujabbarov, B. Ahmedov, \textit{Shadow of rotating non-Kerr black hole}, Phys. Rev. D \textbf{88} (2013), 064004.

\bibitem{I14} F. Atamurotov, A. Abdujabbarov, B. Ahmedov, \textit{Shadow of rotating Hořava-Lifshitz black hole}, Astrophys. Space Sci. \textbf{348} (2013), 179-188.

\bibitem{I15}  A. Abdujabbarov, F. Atamurotov,  Y. Kucukakca, B. Ahmedov, U. Camci, \textit{Shadow of Kerr-Taub-NUT black hole}, Astrophys. Space Sci., \textbf{344} (2013), 429-435.

\bibitem{A1} A. Belhaj,  M. Benali,  Y. Hassouni, \textit{Superentropic Black Hole Shadows in Arbitrary Dimensions}, arXiv preprint arXiv:2203.06774  (2022).

\bibitem{A2} A. Belhaj, H. Belmahi, M. Benali, H. El Moumni, M. A. Essebani, M. B. Sedra, \textit{Optical shadows of rotating Bardeen-AdS black holes}, Mod. Phys. Lett. A (2022), 2250032.

\bibitem{A3} A. Belhaj, M. Benali, H. El Moumni, M. A. Essebani, M. B. Sedra, Y. Sekhmani, \textit{Thermodynamic and Optical Behaviors of Quintessential Hayward-AdS Black Holes}, Inter. Jour. Geom. Meth. Mod. Phys. (2022) 2250096.

\bibitem{A4} A. Belhaj, H. Belmahi, M. Benali, \textit{Superentropic AdS black hole shadows}, Phys. Lett. B \textbf{821} (2021), 136619.

\bibitem{A5} A. Belhaj, H. Belmahi, , M. Benali, W. El Hadri, H. El Moumni, E. Torrente-Lujan,  \textit{Shadows of 5D black holes from string theory}, Phys. Lett. B  \textbf{812} (2021), 136025.

\bibitem{A6}  A. Belhaj, M. Benali, A. El Balali,  W. El Hadri, H. El Moumni, E. Torrente-Lujan, \textit{Black hole shadows in M-theory scenarios}, Inter. Jour. Mod. Phys. D, \textbf{30}(04) (2021), 2150026.

\bibitem{A7} A. Belhaj, M. Benali, A. El Balali, W. El Hadri, H. El Moumni,\textit{ Cosmological constant effect on charged and rotating black hole shadows}, Inter. Jour. Geom. Meth.  Mod. Phys. \textbf{18} (2021) , 2150188.

\bibitem{A8} A. Belhaj, M. Benali, A. El Balali,  H. El Moumni, S. E. Ennadifi,  \textit{Deflection angle and shadow behaviors of quintessential black holes in arbitrary dimensions}, Class.  Quant. Grav. \textbf{37} (2020), 215004.

\bibitem{A9} A. Belhaj, A. El Balali,  W. El Hadri, Y. Hassouni,  E. Torrente-Lujan,  \textit{Phase transition and shadow behaviors of quintessential black holes in M-theory/superstring inspired models}, Inter. Jour. Mod. Phys. A. \textbf{36} (2021), 2150057.

\bibitem{GL1} H. Hoekstra, M. Bartelmann, H. Dahle, H. Israel, M. Limousin, M. Meneghetti, \textit{Masses of galaxy clusters from gravitational lensing}, Space Sci. Rev. \textbf{177} (2013), 75-118. 

\bibitem{GL2} M. M. Brouwer et al, \textit{Studying galaxy troughs and ridges using weak gravitational lensing with the Kilo-Degree Survey}, Mon. Not. Roy. Astron. Soc. \textbf{481} (2018), 5189. 


\bibitem{GL3} F. Bellagamba et al, \textit{AMICO galaxy clusters in KiDS-DR3: weak lensing mass calibration}, Mon. Not. Roy. Astron. Soc. \textbf{484} (2019), 1598.


\bibitem{GL4} R. A. Vanderveld, M. J. Mortonson, W. Hu, T. Eifler, \textit{Testing dark energy paradigms with weak gravitational lensing}, Phys. Rev. D \textbf{85} (2012), 103518.     
\bibitem{GL5} H. J. He and Z. Zhang, \textit{Direct probe of dark energy through gravitational lensing effect}, J. Cosmol. Astropart. Phys. \textbf{2017}, 036.
\bibitem{GL6} S. Cao, G. Covone, Z. H. Zhu, \textit{Testing the dark energy with gravitational lensing statistics}, Astrophys. J. \textbf{755} (2012), 31.
\bibitem{GL7} D. Huterer, D. L. Shafer, \textit{Dark energy two decades after: Observables, probes, consistency tests}, Rep. Prog. Phys. \textbf{81} (2018), 016901.
\bibitem{GL8} S. Jung, C. S. Shin, \textit{Gravitational-wave fringes at LIGO: detecting compact dark matter by gravitational lensing}, Phys. Rev. Lett. \textbf{122} (2019), 041103.
\bibitem{GL9} K. E. Andrade, Q. Minor, A. Nierenberg, M. Kaplinghat, \textit{Detecting dark matter cores in galaxy clusters with strong lensing}, Mon. Not. R. Astron. Soc. \textbf{487} (2019), 1905.
\bibitem{GL10}  B. Turimov, B. Ahmedov, A. Abdujabbarov, C. Bambi, \textit{Gravitational lensing by a magnetized compact object in the presence of plasm}, Int. J. Mod. Phys. D \textbf{28} (2019), 2040013.


\bibitem{GL11} A. Övgün, I. Sakalli, J. Saavedra, \textit{Shadow cast and Deflection angle of Kerr-Newman-Kasuya spacetime}, J. Cosmol. Astropart. Phys. \textbf{2018}, 041. 
\bibitem{GL12} A. Övgün, K. Jusufi,  I. Sakalli, \textit{Gravitational lensing under the effect of Weyl and bumblebee gravities: Applications of Gauss–Bonnet theorem}, Ann. Phys. \textbf{399} (2018), 193. 
\bibitem{GL13} A. Övgün, \textit{Light deflection by Damour-Solodukhin wormholes and Gauss-Bonnet theorem}, Phys. Rev. D \textbf{98} (2018), 044033. 
\bibitem{GL14} A. Övgün, I. Sakalli,   J. Saavedra, \textit{Weak gravitational lensing by Kerr-MOG black hole and Gauss–Bonnet theorem}, Ann. Phys. \textbf{411} (2019), 167978. 
\bibitem{GL15}  K. Jusufi   A. Övgün, \textit{Gravitational lensing by rotating wormholes}, Phys. Rev. D \textbf{97} (2018), 024042. 
\bibitem{GL16} A. Övgün, G. Gyulchev, K. Jusufi, \textit{Weak Gravitational lensing by phantom black holes and phantom wormholes using the Gauss–Bonnet theorem}, Ann. Phys. \textbf{406} (2019), 152. 
\bibitem{GL17} A. Övgün, \textit{Weak field deflection angle by regular black holes with cosmic strings using the Gauss-Bonnet theorem}, Phys. Rev. D \textbf{99} (2019), 104075. 
\bibitem{GL18} K. Jusufi, M. C. Werner, A. Banerjee, A. Övgün, \textit{Light deflection by a rotating global monopole spacetime}, Phys. Rev. D \textbf{95} (2017), 104012. 
\bibitem{GL19} I. Sakalli,   A. Övgün, \textit{Hawking radiation and deflection of light from Rindler modified Schwarzschild black hole}, Europhys. Lett. \textbf{118} (2017), 60006. 
\bibitem{GL20} K. Jusufi, A. Övgün,   A. Banerjee, \textit{Light deflection by charged wormholes in Einstein-Maxwell-dilaton theory}, Phys. Rev. D \textbf{96} (2017), 084036. 
\bibitem{GL21} Y. Kumaran   A. Övgün, \textit{Weak deflection angle of extended uncertainty principle black holes}, Chin. Phys. C \textbf{44} (2020), 025101. 
\bibitem{GL22} K. Jusufi, A. Övgün, J. Saavedra,  Y. Vásquez, P. A. Gonzalez,\textit{ Deflection of light by rotating regular black holes using the Gauss-Bonnet theorem}, Phys. Rev. D \textbf{97} (2018), 124024. 
\bibitem{GL23} W. Javed, J. Abbas,   A. Övgün, \textit{Deflection angle of photon from magnetized black hole and effect of nonlinear electrodynamics}, Eur. Phys. J. C \textbf{79} (2019), 694. 
\bibitem{GL24} K. Jusufi, I. Sakallı,   A. Övgün, \textit{Effect of Lorentz symmetry breaking on the deflection of light in a cosmic string spacetime}, Phys. Rev. D \textbf{96} (2017), 024040. 
\bibitem{GL25} Z. Li,   A. Övgün, \textit{Finite-distance gravitational deflection of massive particles by a Kerr-like black hole in the bumblebee gravity model}, Phys. Rev. D \textbf{101} (2020), 024040. 

  
\bibitem{I16} K. Saurabh, K. Jusufi,  \textit{Imprints of dark matter on black hole shadows using spherical accretions}, Euro. Phys. Jour. C, \textbf{81} (2021), 1-14.

\bibitem{I17}  K. Jusufi, \textit{Quasinormal modes of black holes surrounded by dark matter and their connection with the shadow radius}, Phys. Rev. D, \textbf{101} (2020), 084055.

\bibitem{I18} X. Hou, Z. Xu, M. Zhou, J. Wang, \textit{Black hole shadow of Sgr A* in dark matter halo}, J. Cosmol. Astropart. Phys. \textbf{2018} (07), 015.
\bibitem{I19} T. C. Ma, H. X. Zhang, P. Z. He, H. R. Zhang, Y. Chen, J. B. Deng,  \textit{Shadow cast by a rotating and nonlinear magnetic-charged black hole in perfect fluid dark matter}, Mod. Phys. Lett. A, \textbf{36} (2021), 2150112.
\bibitem{I20} R. C. Pantig, E. T. Rodulfo,   \textit{Rotating dirty black hole and its shadow, Chinese Journal of Physics}, \textbf{68} (2020), 236-257. 
\bibitem{I21} G.S. Bisnovatyi-Kogan, O.Y. Tsupko, \textit{Shadow of a black hole at cosmological distances}, Phys. Rev. D \textbf{98} (2018), 084020.
\bibitem{I22} V. Perlick, O.Y. Tsupko, G.S. Bisnovatyi-Kogan, \textit{Black hole shadow in an expanding universe with a cosmological constant}, Phys. Rev. D \textbf{97} (2018), 104062.





\bibitem{I30} P.V.P. Cunha, C.A.R. Herdeiro, E. Radu, H.F. Runarsson, \textit{Shadows of Kerr black holes with scalar hair}, Phys. Rev. Lett. \textbf{115} (2015), 211102.

\bibitem{I31} X. Hou, Z. Xu, J. Wang, \textit{Rotating black hole shadow in perfect fluid dark matter}, J. Cosmol. Astropart. Phys. \textbf{2018} (12), 040.
\bibitem{I32} S. Haroon, M. Jamil, K. Jusufi, K. Lin, R.B. Mann, \textit{Shadow and deflection angle of rotating black holes in perfect fluid dark matter with a cosmological constant}, Phys. Rev. D \textbf{99} (2019), 044015.


\bibitem{I34} V. C. Rubin, J.W. K. Ford,   N. Thonnard, \textit{ Rotational properties of 21 SC galaxies with a large range of luminosities and radii, from NGC 4605/R= 4kpc/to UGC 2885/R= 122 kpc}, Astrophys. J. \textbf{238} (1980), 471.



\bibitem{I36} K. Akiyama, et al., \textit{Event Horizon Telescope Collaboration}, Astrophys. J. \textbf{875}, L2 (2019), arXiv:1906.11239 [astro-ph.IM]. 
\bibitem{I37} K. Akiyama, et al., \textit{Event Horizon Telescope Collaboration}, Astrophys. J. \textbf{875}, L3 (2019), arXiv:1906.11240 [astro-ph.GA]. 
\bibitem{I38} K. Akiyama, et al., \textit{Event Horizon Telescope Collaboration}, Astrophys. J. \textbf{875}, L4 (2019), arXiv:1906.11241 [astro-ph.GA]. 

\bibitem{I39} K. Akiyama, et al., \textit{Event Horizon Telescope Collaboration}, Astrophys. J. \textbf{875}, L5 (2019), arXiv:1906.11242 [astro-ph.GA]. 

\bibitem{I40} K. Akiyama, et al., \textit{Event Horizon Telescope Collaboration}, Astrophys. J. \textbf{875}, L6 (2019), arXiv:1906.11243 [astro-ph.GA].

\bibitem{I41} B. Toshmatov, Z. Stuchlik,   B. Ahmedov, \textit{Rotating black hole solutions with quintessential energy}, Euro. Phys. Jour. P. \textbf{132} (2017), 98.
\bibitem{I42} S. Shaymatov, B. Ahmedov, Z. Stuchlik,   A. Abdujabbarov, \textit{Effect of an external magnetic field on particle acceleration by a rotating black hole surrounded with quintessential energy}, Intern. Jour.  Mod. Phys. D \textbf{27} (2018), 1850088.

\bibitem{I43} M. H. Li ,  K. C. Yang, \textit{Galactic dark matter in the phantom field}, Phys. Rev. D \textbf{86} (2012), 123015.




\bibitem{I46} A. Das, A. Saha,   S. Gangopadhyay, \textit{Investigation of circular geodesics in a rotating charged black hole in the presence of perfect fluid dark matter}, Class.  Quant. Grav. \textbf{38} (2021), 065015.

\bibitem{II1} A. Övgün, \textit{Weak deflection angle of black-bounce traversable wormholes using Gauss-Bonnet theorem in the dark matter medium}, Turk. J. Phys. \textbf{44} (2020), 465. 

\bibitem{II2} A. Övgün, \textit{Deflection Angle of photons through dark matter by black holes and wormholes using Gauss–Bonnet theorem}, Universe \textbf{5} (2019), 115.

\bibitem{II3} R. C. Pantig,   E. T. Rodulfo, \textit{Weak deflection angle of a dirty black hole}, Chin. J. Phys. \textbf{66} (2020), 691. 
\bibitem{II4} R. C. Pantig,   E. T. Rodulfo, \textit{Rotating dirty black hole and its shadow}, Chin. J. Phys. \textbf{68} (2020), 236.
\bibitem{I47} K. Eda, Y. Itoh, S. Kuroyanagi,   J. Silk, \textit{New probe of dark-matter properties: gravitational waves from an intermediate-mass black hole embedded in a dark-matter minispike}, Phys. Rev. Lett. \textbf{110} (2013), 221101.
\bibitem{I48} C. F. B. Macedo, P. Pani, V. Cardoso,   L. C. B. Crispino, \textit{Into the lair: gravitational-wave signatures of dark matter}, Astrophys. J. \textbf{774} (2013), 48.
\bibitem{I49} E. Barausse, V. Cardoso   P. Pani, \textit{Can environmental effects spoil precision gravitational-wave astrophysics?}, Phys. Rev. D \textbf{89} (2014), 104059.

\bibitem{I50} V. Cardoso,   A. Maselli, \textit{Constraints on the astrophysical environment of binaries with gravitational-wave observations}, Astron. Astrophys. \textbf{644} (2020), A147.
\bibitem{I51} B. J. Kavanagh, D. A. Nichols, G. Bertone   D. D. Gaggero, \textit{Detecting dark matter around black holes with gravitational waves: Effects of dark-matter dynamics on the gravitational waveform}, Phys. Rev. D \textbf{102} (2020), 083006.

\bibitem{Car1}
V.~Cardoso, K.~Destounis, F.~Duque, R.~Panosso Macedo and A.~Maselli, {\it Gravitational Waves from Extreme-Mass-Ratio Systems in Astrophysical Environments}, Phys. Rev. Lett. \textbf{129} 24 (2022)  241103.

\bibitem{Car2}
K.~Destounis, A.~Kulathingal, K.~D.~Kokkotas and G.~O.~Papadopoulos, {\it Gravitational-wave imprints of compact and galactic-scale environments in extreme-mass-ratio binaries}, Phys. Rev. D \textbf{107} 8 (2023)  084027.
\bibitem{Car3}
E.~Figueiredo, A.~Maselli and V.~Cardoso, {\it Black holes surrounded by generic dark matter profiles: Appearance and gravitational-wave emission}, Phys. Rev. D \textbf{107} 10 (2023) 104033.










\bibitem{I52} V. Cardoso, K. Destounis, F. Duque,  R. P. Macedo, A. Maselli, \textit{Black holes in galaxies: Environmental impact on gravitational-wave generation and propagation}, (2022)  Phys. Rev. D \textbf{105}, L061501.

\bibitem{I53} L. Hernquist, \textit{An Analytical Model for Spherical Galaxies and  Bulges}, Astroph. Jour. \textbf{356} (1990), 359.

\bibitem{III1} E. T. Newman, A. I. Janis, \textit{Note on the Kerr spinning-particle metric}, Jour. Math. Phys. \textbf{6} (1965), 915.


\bibitem{III2} S. P. Drake,   P. Szekeres, \textit{Uniqueness of the Newman–Janis algorithm in generating the Kerr–Newman metric}, Gen. Rel. Grav. \textbf{32} (2000), 445-458.

\bibitem{1S1} B. Carter,\textit{ Global structure of the Kerr family of gravitational fields}, Phys. Rev.
\textbf{174},1968, 1559

\bibitem{2S1} S. Vazquez, E. P. Esteban, \textit{Strong field gravitational lensing by a Kerr black hole}, Nuovo Cim.B \textbf{119}, 2004, 489.

\bibitem{GRO} S. E. Motta, T. M. Belloni, L. Stella, T. Muñoz-Darias, R. Fender,  \textit{Precise mass and spin measurements for a stellar-mass black hole through X-ray timing: the case of GRO J1655-40}, Mon. Not. Roy. Astro. Soc. \textbf{437} (2014), p. 2554-2565.

\bibitem{4U} W. R. Morningstar, J. M. Miller, \textit{The spin of the black hole 4U 1543-47}, Astroph. Jour. Lett. \textbf{793} (2014), L33.



\bibitem{deflection} G. Gibbons,   M. Werner, \textit{Applications of the Gauss-Bonnet theorem to gravitational lensing}, Class. Quant. Grav. \textbf{25}, 2008, 235009, arXiv:0807.0854.


\bibitem{A1art} P. Beltracchi, P. Gondolo, \textit{Physical interpretation of Newman-Janis rotating systems. II. General systems},  Phys. Rev. D \textbf{104}, 2021, 124067.
\end{thebibliography}
\end{document}